# Clustering and attention based model for intelligent trading

Mimansa Rana, Nanxiang Mao, Ming Ao, Xiaohui Wu, Poning Liang and Matloob Khushi
The School of Computer Science, The University of Sydney

## ABSTRACT

The foreign exchange market has taken an important role in the global financial market. While foreign exchange trading brings high-yield opportunities to investors, it also brings certain risks. Since the establishment of the foreign exchange market in the 20th century, foreign exchange rate forecasting has become a hot issue studied by scholars from all over the world. Due to the complexity and number of factors affecting the foreign exchange market, technical analysis cannot respond to administrative intervention or unexpected events. Our team chose several pairs of foreign currency historical data and derived technical indicators from 2005 to 2021 as the dataset and established different machine learning models for event driven price prediction for oversold scenario.

# 1. INTRODUCTION

With development of the economy and improvement of people's living standards, the consciousness of investment gradually appears in the minds of the public. The foreign exchange market has far surpassed other markets such as stocks and futures, becoming the largest financial market in the world [1]. The rapid growth in the size of the foreign exchange market has brought high profits and high risks at the same time. So it becomes important to be able to make effective predictions on the trend of foreign exchange prices. Foreign exchange markets change irregularly over time. Under fast changing international pattern and corresponding policies, there are many factors affecting the exchange rate fluctuations. The artificial trading is easy to produce negative emotions and difficult to observe the market situations at all times. On this basis, we adopt machine learning/deep learning models to improve predictive accuracy in foreign exchange market using basic price information as well as derived technical indicators. And eventually, the prediction model can be used by a wide range of stakeholders including banks and hedge funds to diversify their trading strategies as well generate stable profits over time.

# 2. RELATED LITERATURE

## 2.1 Technical indicators

Technical indicators are often used to determine trading strategies within the Forex Market as they can be used to understand trends and signal upcoming changes in market conditions. Below are some indicators that we have found being commonly used during our research.

### 2.1.1 Relative Strength Indicator (RSI)

RSI (relative strength indicator) is used to determine overbought or oversold conditions in the market. It is calculated using the below formula

$$RSI = 100 - \frac{100}{1 + RS} \qquad (1)$$

$$RS = \frac{Avg\ Gain\ over\ N\ days}{Avg\ Loss\ over\ N\ days} \qquad (2)$$

RSI is an oscillator indicator and its value lies between 0 and 100. A value of < 30 indicates that oversold market conditions while an RSI value of >(70-80) indicates overbought market conditions. Oversold market conditions usually means that there is an increase in the likelihood of price going up and an opportunity to buy. Overbought market conditions usually indicate the increase in likelihood of the price going down and an opportunity to sell.

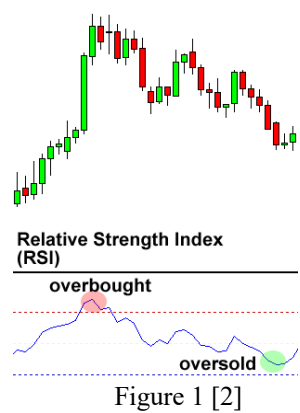
Figure 1 [2]

### 2.1.2 Simple Moving Average (SMA), Exponential Moving Average (EMA), MACD (Moving Average Convergence Divergence)

MACD is a useful indicator to compare short term and long term price trends, which help to determine buy or sell points. In order to calculate MACD we first calculate Simple Moving Average (SMA) and Exponential Moving Average (EMA).

SMA is the average of a range of prices divided by the number of periods in the range.

EMA is similar to SMA except that the more recent prices are given a higher weight during the calculation of the average.

$$MACD = EMA\ (over\ 12\ days) - EMA\ (over\ 26\ days) \qquad (3)$$

The MACD values are usually plotted on the same graph as the 9 day SMA of the forex values (the 9 day SMA line is usually called Signal Line) . The movement of MACD line relative to the Signal line determines if the market is Bullish or Bearish and can be useful to determine if the trader should look for a long or short position.

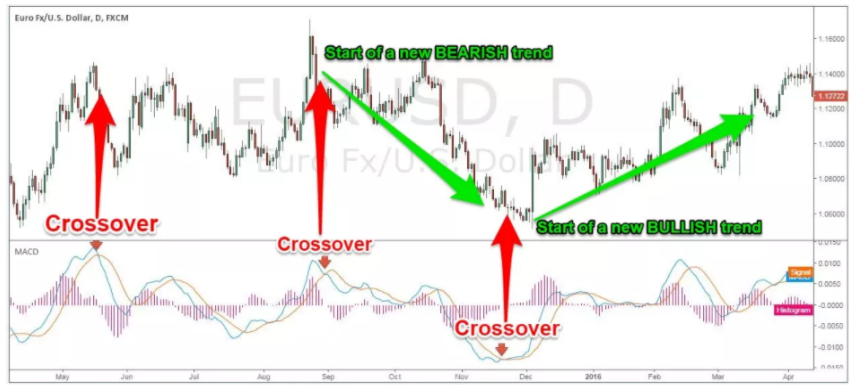

Figure 2 [3]

### 2.1.3 Bollinger Bands

Bollinger Bands are a set of trendlines or bands plotted for a SMA plus/minus 1 standard deviation and 2 standard deviations from SMA. The area represented by 1 SD from SMA on either side represents the Average Band, whereas the area represented by the lines 2 SDs away from SMA indicate the Higher and Lower bands. Approximately 90% of the price movement occurs between the 2 bands (upper and lower), so any changes in price outside the bands could be a significant event. A narrowing or squeezing of the band itself is also seen as a period of low volatility and could signal future increase in volatility.

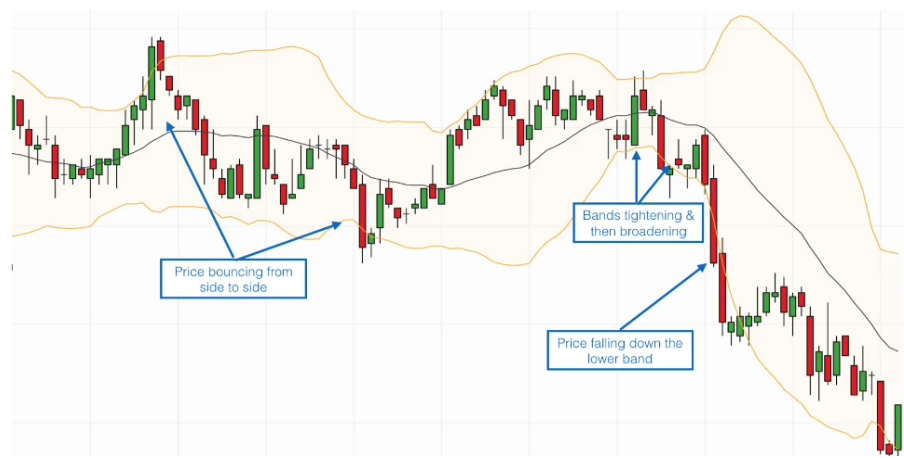

Figure 3 [4]

### 2.1.4 KD – stochastic indicator

Stochastic Oscillaor (%K) is a momentum indicator that is often used to determine signals to a movement in price before it occurs or trend reversals. When depicted on a graph it is usually drawn with %D line which is the 3 period SMA of %K.

% K is calculated using the following formula:

$$\%K = 100 \left[ \frac{C - Ln}{Hn - Ln} \right] \quad (4)$$

Where

    C is current closing price.

    Hn, Ln are the highest and lowest prices respectively in the previous n trading sessions.

The relative movement of the 2 lines (%K and %D) and especially the points when they cross over can be used to predict trend reversals or overbought/oversold conditions.

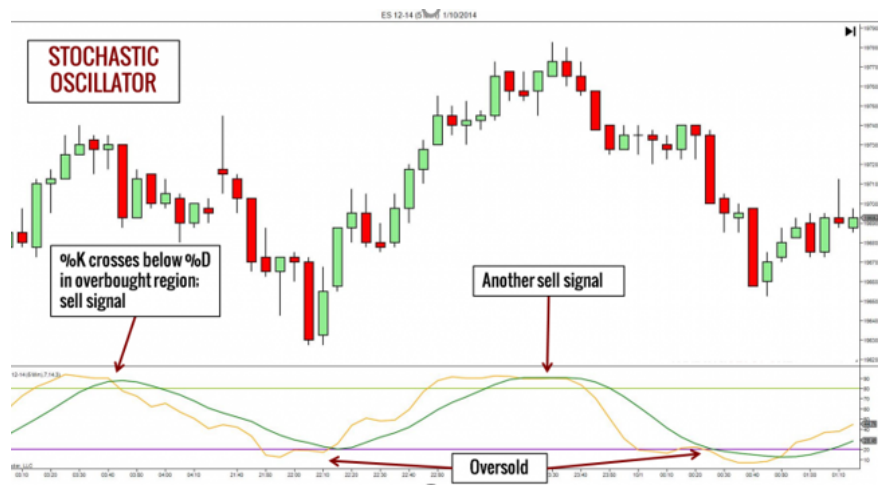

Figure 4 [5]

## 2.1.5 DMI - Directional movement index

Directional movement index is used to identify what direction the asset price is moving. It compares prior high and lows and draw two lines: the positive directional movement line (+DI) and a negative directional movement line (-DI). When +DI is above -DI meaning there is more upward pressure than downward pressure on price.

$$+DI = \left( \frac{\text{Smoothed +DM}}{\text{ATR}} \right) \times 100$$
$$-DI = \left( \frac{\text{Smoothed -DM}}{\text{ATR}} \right) \times 100 \quad (5)$$
$$DX = \left( \frac{|+DI - -DI|}{|+DI + -DI|} \right) \times 100$$

Where:

$$+\text{DM (Directional Movement)} = \text{Current High} - \text{PH}$$
$$\text{PH} = \text{Previous high}$$
$$-\text{DM} = \text{Previous Low} - \text{Current Low}$$
$$\text{Smoothed} +/-\text{DM} = \sum_{t=1}^{14} \text{DM} - \left(\frac{\sum_{t=1}^{14} \text{DM}}{14}\right) + \text{CDM}$$
$$\text{CDM} = \text{Current DM}$$
$$\text{ATR} = \text{Average True Range}$$

### 2.1.6 Concealing Baby Swallow Candlestick pattern

Concealing Baby Swallow Candlestick pattern is made up of four candlesticks. First candlestick: a marubozu candlestick with a bearish body. Second candlestick opens within the prior one's body and close below prior closing price. The third candlestick has long upper shadow and no lower shadow. It opens below prior closing price and upper shadow enters the prior candlestick's body. The fourth one is a tall candlestick with its body covers the pervious candlestick's body and shadows.

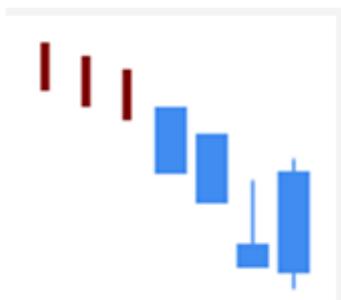 (28)

### 2.1.7 Pattern: Oversold/Overbought

There are two types of oversold scenarios: fundamentally oversold and technically oversold. Fundamentally oversold is when an asset is trading below their true value (overbought is the opposite to oversold). This could be because people have a negative outlook for the asset or any underlying risks that prevent people from buying it. One of popular indicators for a fundamentally oversold is P/E ratio. For technical oversold, it can be defined as a period of time when there has been some significant and consistent downward movement in price without any bounce. And in a technical oversold condition we only consider technical indicators such as RSI and stochastic indicators: an asset is considered to be technically oversold when its 14 periods RSI is equal to or below 30 and overbought when the RSI is equal to or above 80.

## 2.2 Literature Review

The field of machine learning and artificial intelligence is evolving very rapidly and there are new models and architectures being proposed every year that can outperform existing benchmarks. Financial markets specifically are of great interest to researchers, industry and academics as there are significant financial gains at stake. However the challenge with financial market data is that it is noisy, volatile and often unpredictable and there are a few papers we have come across which uses techniques such as PCA[6], wavelet denoising [7],[8], [50] or bagging [9]. [44] employs the use of feature selection to improve performance of stock price prediction.

Several papers deploy architectures based on traditional machine learning models combined with deep learning or neural network architectures to make predictions on future price of stock or forex values based on historical data. [10] proposes a 2-stage approach to predict values of stock market index, a Support Vector Regressor (SVR) based data preparation layer, followed by combinations of Artificial Neural Network (ANN), Random Forest (RF) and SVR. [11] uses a multi-level architecture stacked made up of common machine learning algorithms such as SVM, Random Forest, Dense Layer NN, Naïve Bayes and Bayesian autoregressive trees (BART) and aggregates the scores obtained using a stacked generalisation mechanism. [12] uses Genetic Algorithm along with Neural Network to provide a Forex Trading system. [26] splits a whole time-series data into equal length sub-sequence and use k-means to allocate instances into different groups before applying a LSTM model for training and prediction. [48] uses genetic algorithm to for forex price prediction.

[46] performs a review of the methods used in research papers in the field of forex and stock price prediction and concludes that in recent years the use of deep learning models has increased as these methods provide better results and performance.

Several papers use architectures with neural network-based models that make predictions based on sequential input data (sequence2sequence models) such as [13] which uses an LSTM network to model historical data in order to predict future price movement of stocks to achieve 20% return. LSTM is combined with GRU in [14] to make prediction on FX rates, and the model outperforms a model based on GRU or LSTM alone. [15] uses an LSTM-CNN network to predict future stock price, achieving an annualised return of 33.9% vs 13.5% benchmark. [45] also uses event driven LSTM to predict forex price.

Long short-term memory model has been shown to be very effective in many time-series prediction problems, this model is popular to be used in price prediction in stock market and forex recent years. Since in the forex market, fundamental and technical analysis are two main techniques. [20] used LSTM to forecast directional movement of forex data with technical and macroeconomic indicators. They utilized two different datasets which are macroeconomic data and technical indicator data to train LSTM model, this experiment was found to be quite successful with real world data. For the same data, there are different representations which means it can have various features. [21] developed a long short-term memory-convolutional neural network (LSTM-CNN) model that it can learn different features from a same data, namely, it can combine the features of stock time series and stock chart images to forecast the stock prices. In their experiment, the hybrid model outperforms the single models (CNN and LSTM) with SPDR S&P 500 ETF data. Also, they found that the candlestick chart is the best chart for this hybrid model to forecast stock prices. This research has proved that the combination of neural network models can efficiently reduce the prediction error.

Attention based mechanisms and transformers have gained popularity in the last 1-2 years and they tend to be the State-of-the-Art standard at the moment [16]. These architectures are used in many fields of research from NLP to computer vision. An attention-based LSTM model is used to predict stock price in [18]. The prediction is based on both live Twitter data (using NLP to do sentiment analysis) and technical indicators to achieve an MAE of 0.484 and MSE of 0.627. [19] also uses an attention-based LSTM model to predict stock price. With the help of attention layer, the model is able to learn what part of historical data it needs to put more focus on and it has achieved a better performance than LSTM. The model has achieved a MAPE of 0.00550 vs 0.00838 using LSTM and 0.01863 using ARIMA. [8] proposes first de-noise data (i.e. stock price and volume) using wavelet in order to effectively separate useful signal from noise, and then feed data into an attention based LSTM model that has achieved a MSE of 0.1176 vs 0.1409 from GRU and 0.1839 from LSTM. [47] uses a combination of reinforcement learning and attention with transformer model to make predictions on stock data. Transformer architecture is also used in [43], not just for price prediction but for text mining to create a financial domain specific pretrained sentiment analysis model.

Reinforcement learning includes many different types of algorithms, and these algorithms can make trained models achieve different performances. From conclusion from paper written by Thomas G. Fischer [22], the actor-only approach has a better performance than other methods, however, the critic-only approach are used more widely. And the performance of the critic-actor approach is poor due to the unpreventable correlation generated by the critic-actor approach itself. Besides, the indicators used in the actor-only approach is not decisive since the model can find the policy and inner relation by itself. One of the papers published by Dempster [23] used 14 indicators but they did not help in improving the performance. [49] reviews and critiques existing work that is using reinforcement learning in financial markets.

Stacked autoencoder (SAE) has been widely used in the prediction of time series data due to its excellent performance in depth extraction of data set features [30,31,32,33,34]. The researchers also provided an answer to the question of whether the SAE approach can be applied to financial market forecasting. In 2016, the researchers proposed a prediction model that combined SAE and SVR and made predictions on 28 currency pairs datasets. The ANN and SVR models are used as the benchmark models, and the results show that this model has better performance than ANN and SVR to some extent. For example, the model is 6 times better than the ANN model in MSE [33]. Bao et al. used wavelet transforms (WT), SAE, and LSTM to predict stock prices and compared them with WT+LSTM and LSTM. It is proved that the SAE method can effectively improve the accuracy and profitability of the model prediction [34].

## 3. RESEARCH/PROJECT PROBLEMS

In modern financial market, trading in commodities/shares/foreign currencies/other financial instruments have been heavily adopted with latest technologies. From fast evolving telecommunication technologies to more globalised banking systems, it helps to grow the size of the financial market exponentially. In 2021, average daily turnover for foreign exchange market is over $6.6 trillion [24] vs 2020 USA annual GDP of 20.93 trillion [25]. As a result, people are always looking for new trading strategies using latest technology in order to make profits. Algorithm trading is among one of the most popular area which can be traced back to 1970's when it was first introduced

as "designated order turnaround system" by New York Exchange. Nowadays, algorithm trading has been dominating the space from high frequency trading to portfolio management.

Our project problem is whether we can use the latest deep learning algorithms to improve predictive accuracy in foreign exchange market based on basic price information as well as derived technical indicators. Furthermore, with the help of the deep learning algorithm whether we can develop a profitable trading strategy and also minimise downside risks at the same time.

## 3.1 Research/Project Aims & Objectives

The goal of this project is to adopt various machine learning/deep learning models to make accurate predictions on foreign exchange price movement. Because there are a lot of factors will impact FX market such as political or economic ones that makes it very difficult to predict every move solely based on technical indicators. Therefore, we want to do event driven price predictions on any oversold conditions only. Furthermore, we will also modify some latest deep learning algorithm to ensure its adoption of the task yield even better result.

## 3.2 Research/Project Questions

To achieve a desirable outcome or provide some meaningful insight for other people who will look into this area, there are a few important questions need to answer:

- o  What data or technical indicators are important for the prediction?
- o  What is the ideal sequence length of timestep regarding input data? This will also depend on what algorithm we use since more recent ones (such as attention or transformer) have much better memorability.
- o  How are hyper parameters going to impact our models' performance?
- o  Considering the achievements of different machine learning models in time series forecasting, what type of algorithm are more suitable for this task?
- o  For some algorithms, what types of data pre-processing is required (i.e. PCA, data cleaning and etc)?

- How do we evaluate outcome of our project (i.e. whether we can prove a positive return based on the prediction made by our models or using some traditional metric such as MSE or MAE to measure forecast accuracy)?

### 3.3 Research/Project Scope

The main work of the project was divided into three parts. First, data pre-processing and feature engineering are performed to obtain technical indicators as data features. Then different machine learning models are built to compare their performances. We also try to ensure that each model is to achieve its best possible performance through hyper-parameter tuning.

Data used: several pairs of foreign currency historical data dated from 2005 to 2021 will be used for training and testing for the project.

Success Criteria:

- All team members contribute evenly to our project goal on weekly basis.
- Each person to make significant contribution to developing/implementing machine learning model and fine tuning it in order to optimise its performance.
- Implement models with novel ideas along with outstanding prediction performance.
- The model can be applied to real-world forex forecasting problems with acceptable results.

Potential Risks:

- Timing risk – depending on amount of workload and everyone's capacity there are risks of delivering certain goals/milestones in lateness.
- Technical risk – due to complexity of some models it might be difficult to construct or exceed our current abilities which result in fail in implementation.
- Performance risk - there are chances that our models are unable to generate out-performing results.

# 4. METHODOLOGIES

## 4.1 Methods

### 4.1.1 Clustering + Attention:

Step1 Feature selection:

In total there are more than 130 technical indicators available that can be used as input features of our model. To select the best features, we write a loop function which takes each feature and combine with "Open" and "Close" price to predict high price of next 60 mins. Each feature is trained for 35 epochs and then average loss of the final 10 epochs are calculated and used to rank its importance. We then choose the top 4 most important features along with 'Open" and "Close" as our input features.

Step 2 Clustering:

Clustering is one of the most common machine learning methods to identify subgroups in a dataset so that data in the same subgroups have certain degree of similarities while those in different groups tend to have different patterns. In other words, it helps to discover homogeneous subgroups within a dataset such that data within each cluster are as similar as possible and at the same time keep each cluster as far (different) as possible. In the task of foreign exchange prediction, we use clustering to divide our training dataset into various clusters based on input features (technical indicators) such that data with similar price patterns can be grouped together. As a result, we expect it to segregate instances that are heavily driven by news or any other non-technical factors from those can be solely predictable based on technical indicators. The main clustering method we used is Kmeans: the algorithm randomly initializes k number of centroids and assign each data point to the cluster of the closest centroid. Then for each cluster, it re-calculates centroids by finding mean of the data points previously assigned. The algorithm repeats the above steps by re-assigning all the datapoint to the closest centroid and re-calculating the centroid again and again until there are no further updates/changes to all the centroids. Please see below objective function for Kmeans. Please note that Wik = 1 if data point x belongs to cluster k, otherwise wik = 0.

$$J = \sum_{i=1}^{m} \sum_{k=1}^{K} w_{ik} \|x^i - \mu_k\|^2 \qquad (29)$$

Step 3 Prediction:

We apply BiLSTM with attention to each of the cluster from step 1 and make prediction for the next 60 mins high. Dot product attention was used to calculate attention score based on final hidden state and output states of the BiLSTM. By using attention, the model can eliminate locality bias since it can attend to any information from any time step using the attention layer. There are four major steps involved in attention layer. First, to calculate dot product between last hidden state and output state of the BiLSTM. Secondly, to work out probability distribution based on the output from above using softmax function. Thirdly, to use the probability distribution from previous step to calculate weighted sum of the output states. Lastly, the weight sum and last hidden state will be passed through another RNN before concatenating its output to last hidden states and passing it through a linear layer. Please see below figure 4.1.1.1 for the architecture of the attention model.

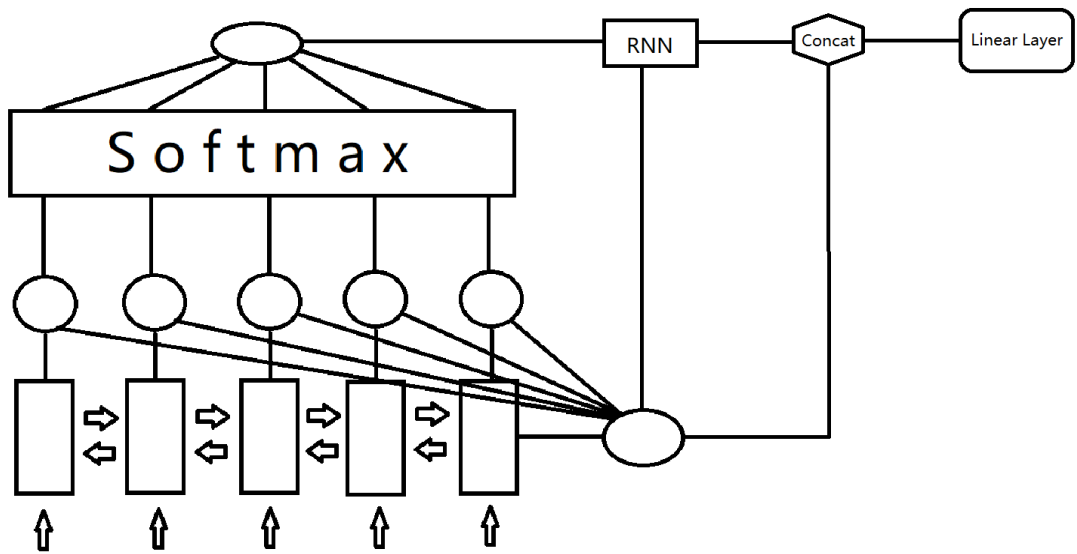

Fig 4.1.1.1

### 4.1.2 Transformer Multi head attention

The components of transformer architecture we use in our model are as per below:

- Multi – head Self - attention

Introduced by [39], this involves taking the attention score of a sequence with respect to the sequence itself to determine which parts of the sequence are the most important. The attention score is calculated using the input sequence as the query, key and value vector. In our model we have calculated attention score using scaled dot product attention. The attention score is calculated n_head times, where n_head is the number of heads of self-attention. The n_head self-attention scores are then concatenated together to form the Multi-Head Attention (MHA) score.

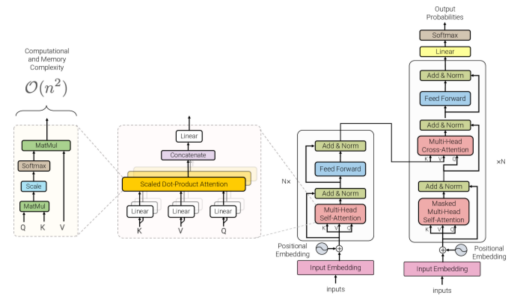

Figure 4.1.2.1 Transformer architecture [41]

The MHA layer forms a part of the Transformer Encoder layer, where the output of the MHA layer is fed into a normalisation layer, followed by a feed forward layer (FF). At each layer normalisation layer, the output of the MHA or FF is added to the original input from the previous layer (a sub layer connection). The encoder layer is stacked 'N' times, with the output of the final layer passed into a linear 'decoder' layer to provide the final prediction.

- Positional embedding/embedding/sequence representation

In transformer architecture there is no inherent notion of sequence input so we need to represent the position of each input relative to the rest of the input space. We have attempted to use 3 methods to do this, with varying results as detailed in section 7 – Results.

1. Time Embedding

Introduced by [40] and based on the implementation by [42], in this technique we use the equation below to calculate the position of our sequence of features relative to other input sequences within the same batch. We calculate the time embedding for the features at each timestep according to the below formula. Each timestep has a linear embedding and a sine embedding and the result is concatenated with the original feature vector.

$$\text{Embedding at } t_i = \begin{cases} w_i * t + bi & if\ i = 0 \\ F(w_i * t + bi) & 1 \leq i \leq k \end{cases} [40]$$

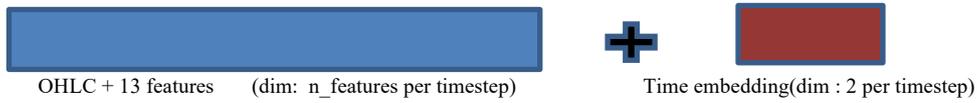

OHLC + 13 features     (dim: n_features per timestep)          Time embedding(dim : 2 per timestep)

Figure 4.1.2.2 Time Embedding

2. Positional Encoding

Positional encoding is a technique used in NLP tasks where the position of each token input is given a position embedding relative to its position in the sequence. For a timestep t and position i within the sequence, each input is given an embedding based on whether it is on an odd or even index within the sequence.

$$\text{Positional Encoding at } p_t^{(i)} = f(t)^i = f(x) = \begin{cases} \sin(\omega_k . t), & if\ i = 2k \\ \cos(\omega_k . t), & if\ i = 2k + 1 \end{cases} [16]$$

$$\text{where } \omega_k = \frac{1}{10000^{\frac{2k}{d}}}$$

3. LSTM

LSTMs process data in a sequential manner so we have attempted to use a single LSTM layer to capture the sequential aspect of our input data. The output of the LSTM layer is fed into the stacked encoders in our experiment.

### 4.1.3 Auto Encoder + CNN + LSTM:

A hybrid model combining stacked autoencoder, CNN and LSTM for forecasting foreign exchange rates is proposed.

Stacked Autoencoder

Manual extraction of features is more difficult in the context of lack of experience in technical forex trading. Autoencoder is a neural network consisting of an input layer, an output layer and a hidden layer, which can automatically extract features from the input data and was first introduced in 1985 [35]. Stacked autoencoder is a simple autoencoder to increase the depth of its hidden layer to obtain better feature extraction ability and training effect. The key idea of autoencoder is to compress the dimensionality of the input data to the number of neurons specified in the encoding layer and then expand to the original dimensionality for reconstruction, forcing the model to extract key features from the data by comparing the output with the initial data.

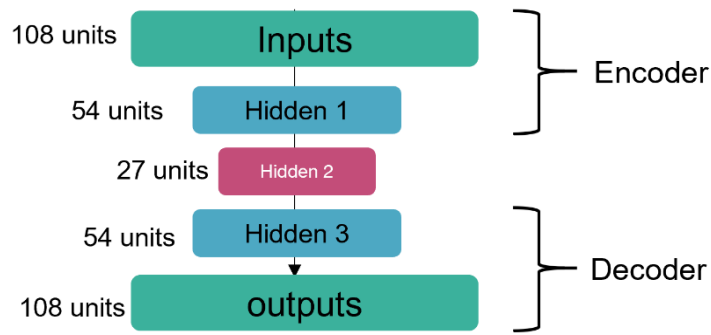

Figure 4.1.3.1 Architecture of SAE

In this project we use a 2-layer stacked autoencoder architecture, and since the stacked autoencoder is symmetric about the hidden layer, 4-layer full architecture is formed, including two encoders and two decoders (see figure 4.1.3.1). The output of the previous hidden layer is used as the input of the next hidden layer. We have one hundred and 8 units in the input encoder layer, and the coding dimension is 27. L1 regularization, is added to reduce the risk of overfitting.

CNN-LSTM model

This model can be divided into four parts, input layer, CNN layers, LSTM layer and dense layers (output layer). The input layer is used to receive the relevant data, then the data will be input into the CNN layers. The CNN layers are consisted of 1D-convolution layer, max pooling layer. The 1D- convolution layer can extract the feature from each sample since all of the features for a point-in-time is in a row. After the convolution is done, the dimension of the data is too large for the convolution kernel, thus, the max pooling layer is carried out to reduce the dimensions of the neural network to avoid overfitting. Then the output of CNN layer will be the input of the LSTM layer. Due to the mechanism of LSTM model which are including forget gate, input gate, output gate and memory cell, it can selectively forget the memory of the input data, thus, the LSTM layer is used for predicting the exact value. The architecture of the CNN-LSTM model is show in figure 4.1.3.2.

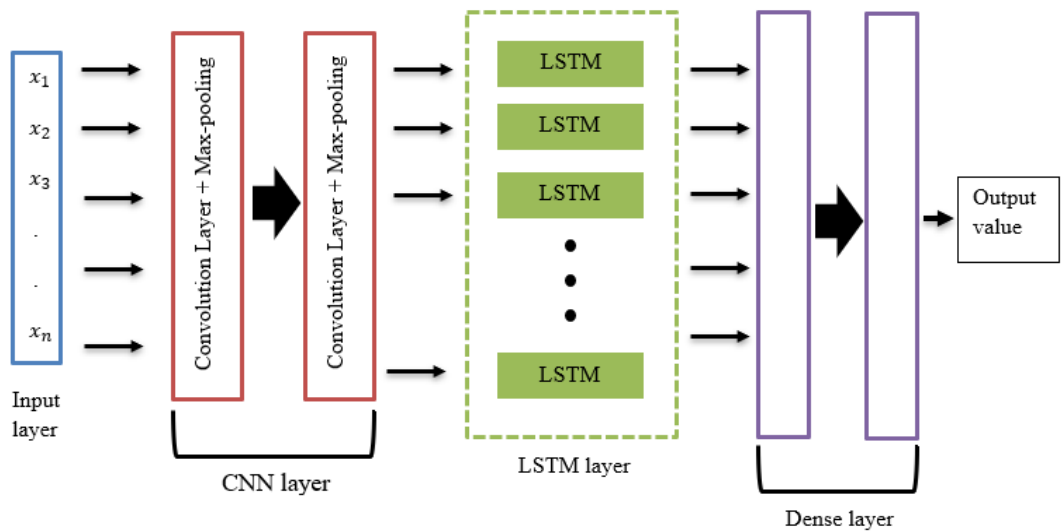
Figure 4.1.3.2: Architecture of CNN-LSTM model

Also, various hyperparameter are involved in the architecture development of this model, through fine tuning these hyperparameters and measuring the accuracy of the validation accuracy, the model has the best performance under the current architecture. The hyperparameters are shown in the Table 4.1.3.3.

Table n: Hyperparameter of CNN-LSTM model

| Hyperparameters | Value |
| --- | --- |
| Convolution Layer Filters | 108 |
| Convolution Layer Kernel size | 2 |
| Convolution Layer Activation Function | sigmoid |
| Convolution Layer kernel initializer | uniform |
| Pooling Layer Pool size | 1 |
| Pooling Layer Padding | valid |
| Number of LSTM Layer Hidden Units | 108 |
| LSTM Layer Activation Function | tanh |
| Time step | 9 |
| Optimizer | adam |
| Loss Function | mse |
| Epochs | 50 |
| Batch size | 60 |

Table 4.1.3.3

### 4.1.4 Reinforcement Learning:

Our aim of using the RL model (reinforcement learning model) is to generate trading strategy. There are many different algorithms for the reinforcement learning, such as Sarsa, deep Q network, and policy gradient. The main differences between all these RL models lie in the methods each model uses to update the policy. However, they all need the same elements including agent, environment, state (observation), action and reward. The agent learns from the environment, executes actions based on the state of the environment (or observations), and try to perform an action which can lead to a higher reward. The algorithm selection and parameter tuning can influence the performance of the model.

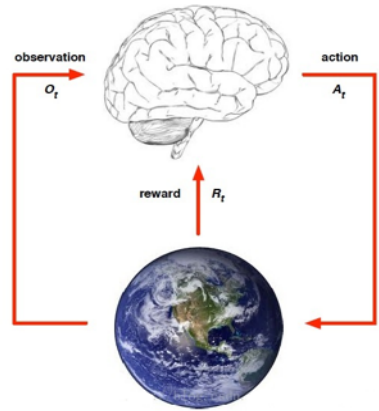

Figure 4.1.4.1. Simple schematic diagram of the reinforcement learning process

1. Action: trading strategy. It is the output of the agent and the input of the environment. In this project, actions are 0 (sell) and 1 (Buy)

2. State (observation) is the output of the environment and the input of agent. In this project, state is the historical data of the foreign exchange market.

3. Reward is the output of the environment and the input of the agent. In this project, reward is the cumulative profit we can obtain from the foreign exchange market.

4. Environment is a place where the agent is trained. In this project, environment is the foreign exchange market.

5. Agent is the brain in Figure 4.1.4.1. It output the action based on the input information and will keep learning until the reward meets the certain requirements.

6. Policy is a mapping of the probability distribution from state to action.

In this foreign exchange marker project, our goal is to identify the right time to buy or sell and make profits from the market. This seems to be suitable for using a value-based algorithm. However, for a value-based model, calculating the Q value and selecting actions based on the Q value will become too complicated for a large amount of data, such as time series data. Secondly, according to the conclusions drawn by Thomas G. Fischer [1], the actor-only approach has a better performance than other methods. Policy is a probabilistic strategy with stronger randomness and flexibility. As a result, we chose policy gradient as our main algorithm for this project.

The objective function is shown below.

$$J(\pi_\theta) = \int_S \rho^\pi(s) \int_a \pi_\theta(s,a) r(s,a) \, da \, ds \qquad (6)$$

Where $\rho^\pi(s)$ is the time spend on the statement s, $\pi_\theta$ is the policy under parameter vector $\theta$, r is the reward.

The policy gradient function is:

$$\nabla_\theta J(\pi_\theta) = \int_S \rho^\mu(s) \int_a \nabla_\theta \mu_\theta(s,a) Q^\mu(s,a) \, da \, ds \qquad (7)$$

where $Q^\mu$ is the expectation of the reward.

We can observe that since we need to integrate two variables at this time, namely action $a$ and statement $s$. It not only needs to consider the state probability distribution, but also the action probability distribution, which increases the computational complexity.

The evaluation of the RL model will be carried out from the following points. The first one is reward curve. By observing whether the curve rises or falls, we will be able to determine whether the model performs good or not. Second, we can decide whether our model can bring us profit by observing the revenue of the RL model on the test dataset.

### 4.2 Data Collection

Data can be collected from various website/sources such as Yahoo financials or Oanda using API. In our case, the OHLC data have been collected and provided to us by our tutor in CSV format.

### 4.3 Data Analysis

We use Talib to calculate a group of technical indicators that can be derived from the OHLC data. Most of our models are mainly focusing on a few different types of technical indicators: moving average which uses long terms and short-term EMA and SMA, MACD, DMI and momentum indicators which uses RSI and stochastic

oscillation. Please see below visual for various technical indicators vs price movement:

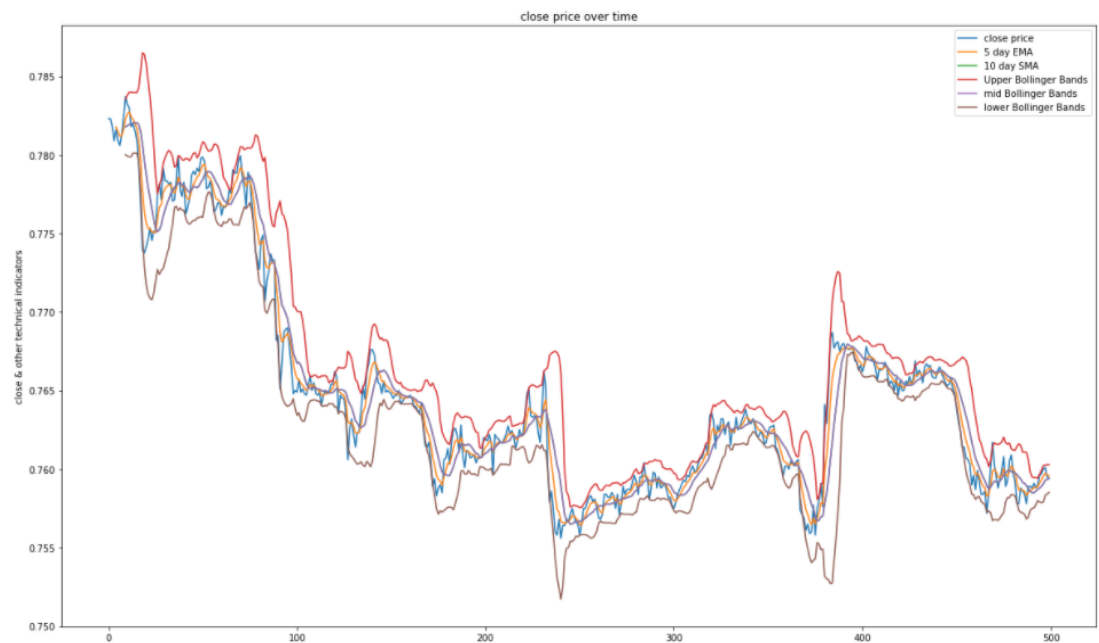

Fig 4.3.1

## 5. RESULTS

### 5.1 Clustering + Attention model:

We have implemented and evaluated the clustering + attention model based on different key parameters including various input sequence length, various future prediction window sizes, different k values in K-means clustering and different cluster threshold in Birch clustering as well as model's performance across different currency pairs. We split dataset from Jan2005 to Jan2020 at a ratio of 80/20 for training and evaluating models' performance (MSE, RMSE and MAE in below tables). We have also heldout one year of data (from Feb2020 to Feb2021) for backtesting purpose only to simulate feeding unseen data into our trained model then to calculate potential profit and loss based on our trading strategy.

In regard to backtesting, we have developed a trading strategy to evaluate how much profit & loss our model can make based on unseen data (Feb20 – Feb21). First, a long position is opened when predicted high price of next hour is higher than current close price (current close price is treated as our entry price in this case). Second, leverage (200x) only applies when the predicted high price is at least 2xMAE higher than current close price. When it comes to exit strategy, there are two scenarios that we exit our long position. First, if the real price reaches the predicted price within next 60 mins then the long position will be closed: profit/loss = predicted high price – entry price - spread. Second, if the real price does not reach our predicted price within next 60 mins, we exit the long position at the end of the 60 mins: profit/loss = closing price in 60 mins – entry price - spread. Here, we use 0.8 bps as the spread between AUD/USD and this spread also include any commissions/brokage fees. Please see below tables for all the experiment results:

| Cluster method | K-means | K-means | K-means | K-means | K-means |
|---|---|---|---|---|---|
| Currency Pair | AUD/USD | AUD/USD | AUD/USD | AUD/USD | AUD/USD |
| Forecast Period | Next 60 mins | Next 60 mins | Next 60 mins | Next 60 mins | Next 60 mins |
| Input Sequence | last 90 mins | last 105 mins | last 120 mins | last 135 mins | last 150 mins |
| # of cluster | 8 | 8 | 8 | 8 | 8 |
| MSE | 5.749E-07 | 5.471E-07 | 6.018E-07 | 4.093E-07 | 7.381E-07 |
| RMSE | 0.0007582 | 0.000739662 | 0.000775758 | 0.0006398 | 0.000859127 |
| MAE | 0.0005069 | 0.000503785 | 0.00054785 | 0.000489 | 0.00055423 |
| Backtest Data Period | 1 year | 1 year | 1 year | 1 year | 1 year |
| Max Leverage Ratio | 200 | 200 | 200 | 200 | 200 |
| Spread | 0.00008 | 0.00008 | 0.00008 | 0.00008 | 0.00008 |
| Backtest P&L with $10k initial capital | **$17,659** | **$11,296** | **$23,699** | **$13,926** | **$4,059** |
| Lowest Capital level based on trades | $8,703 | $9,890 | $9,888 | $7,525 | $7,473 |
| Lowest Capital level based on Minimum price hit | $6,472 | $6,050 | $6,048 | $5,425 | $2,277 |
| Worst Trade | ($820) | ($5,940) | ($5,940) | ($5,940) | ($5,940) |
| Best Trade | $6,237 | $3,553 | $4,641 | $3,934 | $3,017 |

Table 5.1.1 model's performance on next 60 min with various input seq length

| Cluster method | K-means | K-means | K-means | K-means |
|---|---|---|---|---|
| Currency Pair | AUD/USD | AUD/USD | AUD/USD | AUD/USD |
| Forecast Period | **Next 90 mins** | **Next 90 mins** | **Next 90 mins** | **Next 90 mins** |
| Input Sequence | last 90 mins | last 120 mins | last 150 mins | last 180 mins |
| # of cluster | 8 | 8 | 8 | 8 |
| MSE | 1.086E-06 | 8.496E-07 | 1.159E-06 | 1.0547E-06 |
| RMSE | 0.00104211 | 0.000921737 | 0.0010763 | 0.001026986 |
| MAE | 0.0006615 | 0.0006079 | 0.0006716 | 0.0006645 |
| Backtest Data Period | 1 year | 1 year | 1 year | 1 year |
| Max Leverage Ratio | 200 | 200 | 200 | 200 |
| Spread | 0.00008 | 0.00008 | 0.00008 | 0.00008 |
| Backtest P&L with $10k initial capital | **$16,140** | **$31,718** | **$20,419** | **$12,331** |
| Lowest Capital level based on valid trades | $9,947 | $8,372 | $9,946 | $6,437 |
| Lowest Capital level based on Minimum price hit | $9,321 | $6,103 | $6,106 | $3,857 |
| Worst Trade | ($60) | ($3,180) | ($3,180) | ($3,180) |
| Best Trade | $4,340 | $4,012 | $4,484 | $4,041 |

Table 5.1.2 model's performance on next 90 min with various input sequence length

| | | | | |
|---|---|---|---|---|
| Cluster method | K-means | K-means | K-means | K-means |
| Currency Pair | AUD/USD | AUD/USD | AUD/USD | AUD/USD |
| Forecast Period | **Next 120 mins** | **Next 120 mins** | **Next 120 mins** | **Next 120 mins** |
| Input Sequence | last 120 mins | last 150 mins | last 180 mins | last 210 mins |
| # of cluster | 8 | 8 | 8 | 8 |
| MSE | 1.2936E-06 | 1.0893E-06 | 1.2903E-06 | 0.000001221 |
| RMSE | 0.001137365 | 0.001043695 | 0.001135914 | 0.001104989 |
| MAE | 0.000734432 | 0.000676212 | 0.000742177 | 0.000750904 |
| Backtest Data Period | 1 year | 1 year | 1 year | 1 year |
| Max Leverage Ratio | 200 | 200 | 200 | 200 |
| Spread | 0.00008 | 0.00008 | 0.00008 | 0.00008 |
| Backtest P&L with $10k initial capital | **$22,628** | **$32,264** | **$25,786** | **$29,757** |
| Lowest Capital level based on valid trades | $9,448 | $1,598 | $9,795 | $9,898 |
| Lowest Capital level based on Minimum price hit | $8,787 | ($484) | $9,377 | $6,082 |
| Worst Trade | ($3,780) | ($5,120) | ($820) | ($1,600) |
| Best Trade | $4,635 | $4,894 | $6,237 | $6,336 |

Table 5.1.3 model's performance on next 120 min with various input sequence length

| Cluster method | K-means | K-means | K-means | K-means | K-means | K-means |
|---|---|---|---|---|---|---|
| Currency Pair | AUD/USD | AUD/USD | AUD/USD | AUD/USD | AUD/USD | AUD/USD |
| Forecast Period | **next 30 min** | **Next 45 mins** | **next 60 min** | **next 75 min** | **Next 90 mins** | **Next 105 mins** |
| Input Sequence | last 90 mins | last 90 mins | last 90 mins | last 90 mins | last 90 mins | last 90 mins |
| # of cluster | 8 | 8 | 8 | 8 | 8 | 8 |
| MSE | 3.648E-07 | 4.292E-07 | 5.749E-07 | 7.98E-07 | 0.000001086 | 1.2771E-06 |
| RMSE | 0.000603987 | 0.000655134 | 0.000758222 | 0.0008933 | 0.001042113 | 0.00113009 |
| MAE | 0.00039361 | 0.000448492 | 0.000506943 | 0.000577 | 0.0006615 | 0.000717 |
| Backtest Data Period | 1 year | 1 year | 1 year | 1 year | 1 year | 1 year |
| Max Leverage Ratio | 200 | 200 | 200 | 200 | 200 | 200 |
| Spread | 0.00008 | 0.00008 | 0.00008 | 0.00008 | 0.00008 | 0.00008 |
| Backtest P&L with $10k initial capital | **$3,750** | **$9,951** | **$17,659** | **$13,955** | **$16,140** | **$16,953** |
| Lowest Capital level based on valid trades | $9,396 | $9,881 | $8,703 | $8,910 | $9,947 | $7,296 |
| Lowest Capital level based on Minimum price hit | $8,396 | $7,617 | $6,472 | $6,816 | $9,321 | $5,210 |
| Worst Trade | ($7,220) | ($4,340) | ($820) | ($2,100) | ($60) | ($3,240) |
| Best Trade | $2,258 | $2,545 | $6,237 | $3,090 | $4,340 | $4,439 |

Table 5.1.4 model's performance on various prediction window size using last 90 mins data

| Cluster method | K-means | K-means | K-means | K-means |
|---|---|---|---|---|
| Currency Pair | AUD/USD | CAD/USD | NZD/USD | CHF/USD |
| Forecast Period | **Next 60 mins** | **Next 60 mins** | **Next 60 mins** | **Next 60 mins** |
| Input Sequence | last 135 mins | last 135 mins | last 135 mins | last 135 mins |
| # of cluster | 8 | 8 | 8 | 8 |
| MSE | 4.093E-07 | 1.0859E-06 | 6.48E-07 | 0.000001084 |
| RMSE | 0.00063977 | 0.001042065 | 0.000805 | 0.001041153 |
| MAE | 0.00048903 | 0.00066738 | 0.000587 | 0.0007442 |
| Backtest Data Period | 1 year | 1 year | 1 year | 1 year |
| Max Leverage Ratio | 200 | 200 | 200 | 200 |
| Spread | 0.00008 | 0.0002 | 0.0002 | 0.00017 |
| Backtest P&L with $10k initial capital | **$13,926** | **$14,371** | **$2,050** | **$19,308** |
| Lowest Capital level based on valid trades | $7,525 | $9,755 | $5,721 | $1,072 |
| Lowest Capital level based on Minimum price hit | $5,425 | $7,648 | $607 | ($1,188) |
| Worst Trade | ($5,940) | ($5,700) | ($6,060) | ($8,340) |
| Best Trade | $3,934 | $5,742 | $3,386 | $5,113 |

Table 5.1.5 model's performance on different currency pairs

| Cluster method | K-means | K-means | K-means | K-means |
|---|---|---|---|---|
| Currency Pair | JPY/USD | EUR/USD | GBP/USD | JPY/GBP |
| Forecast Period | **Next 60 mins** | **Next 60 mins** | **Next 60 mins** | **Next 60 mins** |
| Input Sequence | last 135 mins | last 135 mins | last 135 mins | last 135 mins |
| # of cluster | 8 | 8 | 8 | 8 |
| MSE | 1.5953 | 1.1539E-06 | 3.5705E-06 | 10.81685 |
| RMSE | 1.26305186 | 0.0010742 | 0.001889577 | 3.288897992 |
| Test MAE +1 epoch vs above | 1.0716 | 0.00075566 | 0.00101 | 2.7524 |
| Backtest Data Period | 1 year | 1 year | 1 year | 1 year |
| Max Leverage Ratio | 200 | 200 | 200 | 200 |
| Spread | 0.01 | 0.00008 | 0.00013 | 0.02 |
| Backtest P&L with $10k initial capital | **($1,211,756)** | **$3,635** | **$13,748** | **($924,372)** |
| Lowest Capital level based on valid trades | ($1,484,846) | $9,999 | $9,141 | ($2,331,331) |
| Lowest Capital level based on Minimum price hit | ($1,594,846) | $9,037 | $2,254 | ($4,188,441) |
| Worst Trade | ($672,000) | ($32) | ($11,060) | ($2,912,000) |
| Best Trade | $912,000 | $3,645 | $6,775 | $744,000 |

Table 5.1.6 model's performance on different currency pairs

| Cluster method | Birch | Birch | Birch | Birch | Birch | Birch |
|---|---|---|---|---|---|---|
| Currency Pair | AUD/USD | AUD/USD | AUD/USD | AUD/USD | AUD/USD | AUD/USD |
| Cluster Threshold | 0.01 | 0.05 | 0.1 | 0.2 | 0.3 | 0.5 |
| Forecast Period | **Next 60 mins** | **Next 60 mins** | **Next 60 mins** | **Next 60 mins** | **Next 60 mins** | **Next 60 mins** |
| Input Sequence | last 135 mins | last 135 mins | last 135 mins | last 135 mins | last 135 mins | last 135 mins |
| # of cluster | 8 | 8 | 8 | 8 but system only generate 5 | 8 but system only generate 5 | 8 but system only generate 3 |
| MSE | 6.46E-07 | 6.917E-07 | 5.123E-07 | 5.205E-07 | 8.336E-07 | 9.414E-07 |
| RMSE | 0.000803 | 0.00083169 | 0.00071575 | 0.00072146 | 0.00091302 | 0.0009703 |
| MAE | 0.000522 | 0.00054037 | 0.0005196 | 0.0004985 | 0.0005616 | 0.0005952 |
| Backtest Data Period | 1 year | 1 year | 1 year | 1 year | 1 year | 1 year |

| Max Leverage Ratio | 200 | 200 | 200 | 200 | 200 | 200 |
|---|---|---|---|---|---|---|
| Spread | 0.00008 | 0.00008 | 0.00008 | 0.00008 | 0.00008 | 0.00008 |
| Backtest P&L with $10k initial capital | **$15,228** | **$17,034** | **$12,028** | **$5,401** | **$7,198** | **$9,522** |
| Lowest Capital level based on valid trades | $9,995 | $9,992 | $5,296 | $7,281 | $2,753 | ($75,912) |
| Lowest Capital level based on Minimum price hit | $8,506 | $8,396 | ($1,584) | $5,185 | ($1,387) | ($89,091) |
| Worst Trade | ($120) | ($5,940) | ($5,940) | ($2,180) | ($5,940) | ($29,840) |
| Best Trade | $3,175 | $3,436 | $3,234 | $2,881 | $3,123 | $15,918 |

Table 5.1.7 model's performance using Birch cluster (k=8) with different Cluster Threshold value

| Cluster method | Birch | Birch | Birch | Birch |
|---|---|---|---|---|
| Currency Pair | AUD/USD | AUD/USD | AUD/USD | AUD/USD |
| Cluster Threshold | 0.01 | 0.05 | 0.1 | 0.2 |
| Forecast Period | **Next 60 mins** | **Next 60 mins** | **Next 60 mins** | **Next 60 mins** |
| Input Sequence | last 135 mins | last 135 mins | last 135 mins | last 135 mins |
| # of cluster | 10 | 10 | 10 | 10 but system reduces to 9 |
| MSE | 1.24E-06 | 8.128E-07 | 8.571E-07 | 6.195E-07 |
| RMSE | 0.001112 | 0.00090155 | 0.0009258 | 0.000787083 |
| MAE | 0.000564 | 0.0005645 | 0.0005769 | 0.0005149 |
| Backtest Data Period | 1 year | 1 year | 1 year | 1 year |
| Max Leverage Ratio | 200 | 200 | 200 | 200 |
| Spread | 0.00008 | 0.00008 | 0.00008 | 0.00008 |
| Backtest P&L with $10k initial capital | **$20,824** | **$23,267** | **$7,765** | **$5,115** |
| Lowest Capital level based on valid trades | $9,910 | $9,986 | $9,827 | $9,600 |
| Lowest Capital level based on Minimum price hit | $8,351 | $8,435 | $6,835 | $5,791 |
| Worst Trade | ($800) | ($61) | ($5,940) | ($5,940) |
| Best Trade | $6,628 | $6,353 | $3,761 | $2,767 |

Table 5.1.8 model's performance using Birch cluster (k=10) with different Cluster Threshold value

| Cluster method | K-means | K-means | K-means | K-means | K-means | K-means |
|---|---|---|---|---|---|---|
| Currency Pair | AUD/USD | AUD/USD | AUD/USD | AUD/USD | AUD/USD | AUD/USD |
| Forecast Period | Next 60 mins | Next 60 mins | Next 60 mins | Next 60 mins | Next 60 mins | Next 60 mins |
| Input Sequence | last 135 mins | last 135 mins | last 135 mins | last 135 mins | last 135 mins | last 135 mins |
| # of cluster | 1 | 2 | 3 | 4 | 5 | 6 |
| MSE | 1.02E-06 | 8.54E-07 | 1.11E-06 | 6.84E-07 | 6.84E-07 | 5.18E-07 |
| RMSE | 0.001008 | 0.000924 | 0.001052 | 0.000827 | 0.000827 | 0.00072 |
| MAE | 0.00066 | 0.000607 | 0.000585 | 0.000559 | 0.00054 | 0.000513 |
| Backtest Data Period | 1 year | 1 year | 1 year | 1 year | 1 year | 1 year |
| Max Leverage Ratio | 200 | 200 | 200 | 200 | 200 | 200 |
| Spread | 0.00008 | 0.00008 | 0.00008 | 0.00008 | 0.00008 | 0.00008 |
| Backtest P&L with $10k initial capital | **($8,082)** | **$29,094** | **$15,143** | **$12,842** | **$1,295** | **$20,661** |
| Lowest Capital level based on valid trades | ($127,510) | ($49,644) | ($52,152) | $5,721 | $2,097 | $9,995 |
| Lowest Capital level based on Minimum price hit | ($140,690) | ($62,824) | ($65,332) | $1,581 | ($847) | $8,374 |
| Worst Trade | ($29,840) | ($29,840) | ($29,840) | ($5,940) | ($5,940) | ($5,940) |
| Best Trade | $14,227 | $15,141 | $15,009 | $3,280 | $3,088 | $3,400 |

Table 5.1.9 model's performance using k-means cluster with different k value

| Cluster method | K-means | K-means | K-means | K-means | K-means | K-means |
|---|---|---|---|---|---|---|
| Currency Pair | AUD/USD | AUD/USD | AUD/USD | AUD/USD | AUD/USD | AUD/USD |
| Forecast Period | Next 60 mins | Next 60 mins | Next 60 mins | Next 60 mins | Next 60 mins | Next 60 mins |
| Input Sequence | last 135 mins | last 135 mins | last 135 mins | last 135 mins | last 135 mins | last 135 mins |
| # of cluster | 7 | 8 | 9 | 10 | 11 | 12 |
| MSE | 7.5E-07 | 4.093E-07 | 4.26E-07 | 4.63E-07 | 4.78E-07 | 4.76E-07 |
| RMSE | 0.000866 | 0.0006397 | 0.000653 | 0.000681 | 0.000691 | 0.00069 |
| MAE | 0.000542 | 0.000489 | 0.000474 | 0.000509 | 0.000495 | 0.000475 |
| Backtest Data Period | 1 year | 1 year | 1 year | 1 year | 1 year | 1 year |
| Max Leverage Ratio | 200 | 200 | 200 | 200 | 200 | 200 |
| Spread | 0.00008 | 0.00008 | 0.00008 | 0.00008 | 0.00008 | 0.00008 |
| Backtest P&L with $10k initial capital | **$5,983** | **$13,926** | **$7,317** | **$7,153** | **$6,034** | **$2,331** |
| Lowest Capital level based on valid trades | $9,998 | $7,525 | $4,474 | $1,911 | $9,878 | $9,895 |
| Lowest Capital level based on Minimum price | $8,371 | $5,425 | $127 | ($2,809) | $6,038 | $7,935 |
| Worst Trade | ($5,940) | ($5,940) | ($5,440) | ($5,440) | ($62) | ($2,160) |
| Best Trade | $2,723 | $3,937 | $3,253 | $3,454 | $3,149 | $2,434 |

Table 5.1.10 model's performance using k-means cluster with different K value

| Cluster method | K-means | K-means | K-means | K-means | K-means | K-means |
|---|---|---|---|---|---|---|
| Currency Pair | AUD/USD | AUD/USD | AUD/USD | AUD/USD | AUD/USD | AUD/USD |
| Forecast Period | Next 60 mins | Next 60 mins | Next 60 mins | Next 60 mins | Next 60 mins | Next 60 mins |
| Input Sequence | last 135 mins | last 135 mins | last 135 mins | last 135 mins | last 135 mins | last 135 mins |
| # of cluster | 13 | 14 | 15 | 16 | 17 | 18 |
| MSE | 6.78E-07 | 4.91E-07 | 4.52E-07 | 8.6E-07 | 4E-07 | 4.25E-07 |
| RMSE | 0.000824 | 0.000701 | 0.000673 | 0.000927 | 0.000632 | 0.000652 |
| MAE | 0.000638 | 0.000542 | 0.000483 | 0.000791 | 0.000468 | 0.000472 |
| Backtest Data Period | 1 year | 1 year | 1 year | 1 year | 1 year | 1 year |
| Max Leverage Ratio | 200 | 200 | 200 | 200 | 200 | 200 |
| Spread | 0.00008 | 0.00008 | 0.00008 | 0.00008 | 0.00008 | 0.00008 |
| Backtest P&L with $10k initial capital | **$5,354** | **$7,297** | **$2,310** | **$2,235** | **$2,301** | **$310** |
| Lowest Capital level based on valid trades | $9,905 | $9,998 | $9,910 | $4,486 | $1,885 | $4,487 |
| Lowest Capital level based on Minimum price hit | $6,065 | $6,486 | $7,881 | $139 | ($2,835) | ($140) |
| Worst Trade | ($2,160) | ($2,160) | ($2,160) | ($5,440) | ($5,440) | ($5,440) |
| Best Trade | $3,348 | $2,984 | $2,340 | $4,712 | $3,268 | $3,082 |

Table 5.1.11 model's performance using k-means cluster with different K value

| Cluster method | K-means | K-means | K-means | K-means |
|---|---|---|---|---|
| Currency Pair | AUD/USD | AUD/USD | AUD/USD | AUD/USD |
| Forecast Period | **Next 60 mins** | **Next 60 mins** | **Next 60 mins** | **Next 60 mins** |
| Input Sequence | last 135 mins | last 135 mins | last 135 mins | last 135 mins |
| Percentage | 20% | 13% | 12% | 11% |
| Accumulative | 20% | 34% | 45% | 57% |
| Cluster Size | 3297 | 2156 | 1866 | 1851 |
| Cluster # | 1st | 2nd | 3rd | 4th |
| MSE | 4.954E-07 | 1.19E-06 | 8.01E-07 | 1.07E-06 |
| RMSE | 0.000703847 | 0.001092 | 0.000895 | 0.001035 |
| MAE | 0.0005075 | 0.000609 | 0.000687 | 0.000743 |

Table 5.1.12 performance by cluster with k-mean (k=9)

| Cluster method | K-means | K-means | K-means | K-means | K-means |
|---|---|---|---|---|---|
| Currency Pair | AUD/USD | AUD/USD | AUD/USD | AUD/USD | AUD/USD |
| Forecast Period | **Next 60 mins** | **Next 60 mins** | **Next 60 mins** | **Next 60 mins** | **Next 60 mins** |
| Input Sequence | last 135 mins | last 135 mins | last 135 mins | last 135 mins | last 135 mins |
| Percentage | 11% | 11% | 7% | 7% | 6% |
| Accumulative | 68% | 79% | 86% | 94% | 100% |
| Cluster Size | 1844 | 1741 | 1201 | 1165 | 1042 |

| Cluster # | 5th | 6th | 7th | 8th | 9th |
|---|---|---|---|---|---|
| MSE | 1.18E-06 | 1.35E-06 | 2.28E-06 | 2.94E-06 | 6.41E-06 |
| RMSE | 0.001086 | 0.00116 | 0.00151 | 0.001713 | 0.002531 |
| MAE | 0.000723 | 0.000755 | 0.000788 | 0.00122 | 0.002236 |

Table 5.1.13 performance by cluster with k-mean (k=9)

Final result on test data:

In this scenario, we split the 15 mins AUD/USD dataset to training (02Jan2005-02Apr2019) and testing (03Apr2019 – 28Feb2021) right at the beginning before any feature generation or shuffling. We also keep 24 hours gap between training and testing data just to ensure there is zero information leakage from testing to training (since all features are calculated based on information that are no longer than past 4 hours). Then we generate features for training and testing separately and all models including clustering was trained based on training data only. Please see below model's performance based on test data:

| Cluster method | Kmeans |
|---|---|
| Currency Pair | AUD/USD |
| Forecast Period | next 60 mins |
| Input Feature | last 135 mins |
| # of cluster | 8 |
| Test MSE | 0.00000063 |
| Test RMSE | 0.000793725 |
| Test MAE +1 epoch vs above | 0.00061 |
| Backtest Data Period | 1 year and 11 months |
| Max Leverage Ratio | 200 |
| Spread | 0.00008 |
| Backtest P&L with $10k initial capital | **$24,709** |
| Lowest Capital level based on valid trades | $9,990 |
| Lowest Capital level based on Minimum price hit | $6,475 |
| Worst Trade | ($5,939) |
| Best Trade | $3,988 |

Table 5.1.14

## 5.2 Transformer Multi Head Attention

The data is split up as 60% training, 20% validation and 20% test. All results are based on predictions on test data. All outputs are after training for 20 epochs and with learning rate of 0.1 using SGD optimizer and a batch size of 6 is used during training.

| Method | Best MSE | Best RMSE | Best MAE |
|---|---|---|---|
| Time embedding + Transformer Encoder | 0.000416866 | 0.020417293 | 0.00901897 |
| Positional encoding + Transformer Encoder | 1.679615 | 1.29599 | 1.24654 |
| Time embedding+positional encoding + Transformer encoder | 1.796017 | 1.34015 | 1.29238 |
| LSTM + Transformer Encoder | 2.290292 | 1.5133 | 1.47123 |
| LSTM + Time Embedding +Transformer Encoder | 0.0079 | 0.089 | 0.066 |
| Transformer Encoder only | 0.002223 | 0.047155 | 0.04386 |

The following results are from using time embedding and transformer encoder.

| Number of encoder layers | Test MSE | Test RMSE | Test MAE |
|---|---|---|---|
| 1 | 0.00107 | 0.03273 | 0.01824 |
| 2 | 0.00077 | 0.02786 | 0.014399 |
| 3 | 0.00055 | 0.02365 | 0.007808 |
| 4 | 0.00081 | 0.02847 | 0.009055 |
| 5 | 0.00062 | 0.02494 | 0.019219 |
| 6 | 0.00058 | 0.02412 | 0.00875 |

Table 5.2.2: 4 heads = 4, d_k, d_v = 10, ff_dim = 32

| Parameters | Test MSE | Test RMSE | Test MAE |
|---|---|---|---|
| n_head = 1, d_k, d_v = 10 | 0.02236 | 0.14953 | 0.08529 |
| n_head = 2, d_k, d_v = 10 | 0.000806 | 0.02839 | 0.01244 |
| n_head = 4, d_k, d_v = 10 | 0.00055 | 0.02365 | 0.007808 |
| n_head = 8, d_k, d_v = 10 | 0.000416866 | 0.020417293 | 0.00901897 |
| n_head = 4 d_k, d_v = 4 | 0.000437085 | 0.02090659 | 0.00732209 |
| n_head = 4 d_k, d_v = 19 | 0.00047899 | 0.021885 | 0.0118105 |
| n_head = 8 d_k, d_v = 25 | 0.0009038 | 0.0300633 | 0.0162498 |
| n_head = 4, d_k = 25, d_v = 35 | 0.00055879 | 0.023638 | 0.008724 |

Table 5.2.3 : 3 encoders, ff_dim = 32

| Parameters | Test MSE | Test RMSE | Test MAE |
|---|---|---|---|
| Conv1d FFN d_ff = 128 | 0.000557 | 0.023613 | 0.00785 |
| Conv1d FFN d_ff = 64 | 0.00045212 | 0.02126329 | 0.008011466 |
| Conv1d FFN d_ff = 32 | 0.00055 | 0.02365 | 0.007808 |
| Conv1d FFN d_ff = 16 | 0.00644 | 0.08025 | 0.04356 |
| Conv1d FFN d_ff = 8 | 0.0004727 | 0.021743 | 0.01574 |
| Linear FFN d_ff = 128 | 2.29029 | 1.51337 | 1.47123 |
| Linear FFN d_ff = 8 | 2.319159 | 1.522878 | 1.48101 |

Table 5.2.4 : 3 encoders, 4 heads, d_k, d_v =10

| Currency Pair | MSE | RMSE | MAE |
|---|---|---|---|
| AUD/USD | 0.000416866 | 0.020417293 | 0.00901897 |
| EUR/USD | 0.09946 | 0.31537 | 0.2159 |
| NZD/USD | $4.03248 \times 10^{-5}$ | 0.00635 | 0.004527 |
| CAD/USD | 0.17537 | 0.41878 | 0.28885 |
| GBP/USD | 0.13925 | 0.37316 | 0.31377 |
| CHF/USD | 0.12856 | 0.35855 | 0.29731 |
| JPY/USD | 0.5648103 | 0.75153 | 0.726237 |
| EUR/AUD | 0.21321 | 0.461748 | 0.38735 |

Table 5.2.5 : 3 encoders, 4 heads, d_k, d_v = 10, ff_dim = 32

## 5.3 Auto Encoder + CNN + LSTM:

To avoid data leakage and accurately simulate the real prediction environment. We use the last year of data as the test set for back testing. The first 80% of the remaining data is used as the training set and the last 20% as the validation set. The hybrid model combined with CNN and LSTM was used as the benchmark model, and the prediction performance of the model was evaluated by comparing the MAE, MSE and RMSE results of the model on the validation set. Figure 5.3.1 shows the predicted data and the corresponding actual data for two models using AUD/USD 15min data to predict the maximum price for the next hour. Figure 5.3.2 shows the results of running our model on the main 8 currency pairs.

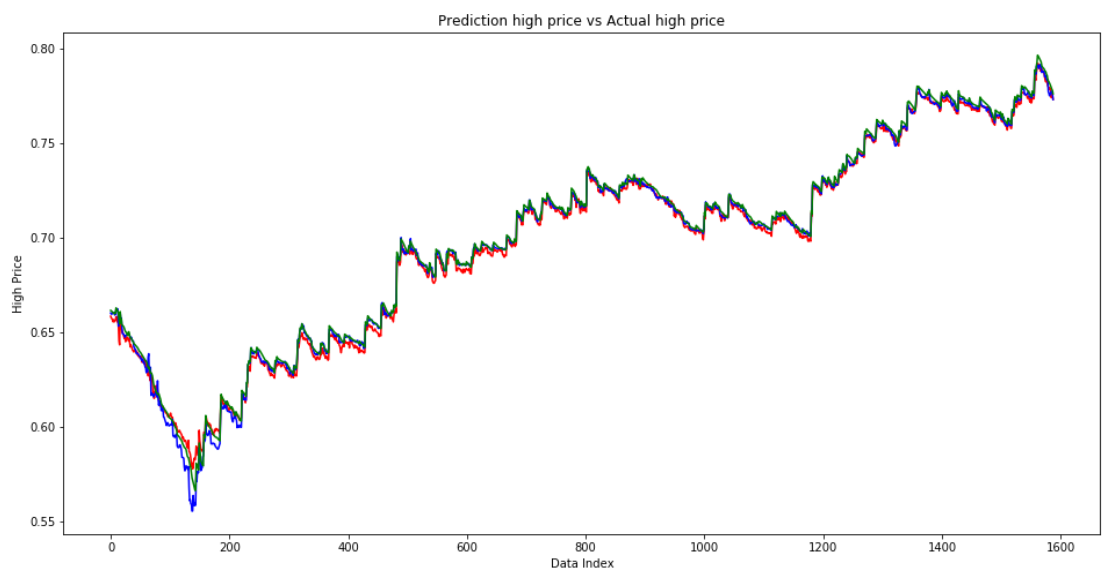

Figure 5.3.1 Actual(blue) and predicted from SAE+CNN+LSTM (red), CNN+LSTM (green) model for AUD/USD.

| Currency Pair | AUD/USD | | EUR/USD | |
|---|---|---|---|---|
| Method | SAE+CNN+LSTM | CNN+LSTM | SAE+CNN+LSTM | CNN+LSTM |
| MAE | 0.044161 | 0.0441207 | 0.053629 | 0.0541307 |
| MSE | 0.002936 | 0.0029309 | 0.004467 | 0.0045622 |
| RMSE | 0.054184 | 0.0541378 | 0.066832 | 0.067447954 |
| Backtest Data Period | 1 year | 1 year | 1 year | 1 year |
| Initial Capital | 10000 | 10000 | 10000 | 10000 |
| Max Leverage Ratio | 200 | 200 | 200 | 200 |
| Profit | 215.178 | -21.96859 | 110.519900 | -178.3685 |

| Currency pair | GBP/JPY | | USD/CAD | |
|---|---|---|---|---|
| Method | SAE+CNN+LSTM | CNN+LSTM | SAE+CNN+LSTM | CNN+LSTM |
| MAE | 23.039691 | 6.316576 | 0.03459609 | 0.035410936 |
| MSE | 562.825276 | 64.2972054 | 0.001968746 | 0.002049775 |
| RMSE | 23.723939 | 8.018553822 | 0.044370559 | 0.045274442 |
| Backtest Data Period | 1 year | 1 year | 1 year | 1 year |
| Initial Capital | 10000 | 10000 | 10000 | 10000 |
| Max Leverage Ratio | 200 | 200 | 200 | 200 |
| Profit | 216619.600000 | 57964.54551 | -1598.313347 | -57.1 |

| Currency pair | GBP/USD | | NZD/USD | |
|---|---|---|---|---|
| Method | SAE+CNN+LSTM | CNN+LSTM | SAE+CNN+LSTM | CNN+LSTM |
| MAE | 0.052105 | 0.0525298 | 0.035086 | 0.035007937 |
| MSE | 0.004426 | 0.0044909 | 0.001862 | 0.0018537 |
| RMSE | 0.066529 | 0.006701399 | 0.043156 | 0.043055629 |
| Backtest Data Period | 1 year | 1 year | 1 year | 1 year |
| Initial Capital | 10000 | 10000 | 10000 | 10000 |
| Max Leverage Ratio | 200 | 200 | 200 | 200 |
| Profit | -1677.067834 | -2649.6711 | -5356.270370 | -5515.332 |

| Currency pair | USD/CHF | | USD/JPY | |
|---|---|---|---|---|
| Method | SAE+CNN+LSTM | CNN+LSTM | SAE+CNN+LSTM | CNN+LSTM |
| MAE | 0.020845 | 0.020592917 | 9.624958 | 2.760108891 |
| MSE | 0.00070155 | 0.000683679 | 98.547680 | 12.03439025 |
| RMSE | 0.026486786 | 0.026147262 | 9.927118 | 3.469061869 |
| Backtest Data Period | 1 year | 1 year | 1 year | 1 year |
| Initial Capital | 10000 | 10000 | 10000 | 10000 |
| Max Leverage Ratio | 200 | 200 | 200 | 200 |
| Profit | -219.5553085 | 129.7413082 | 0.000000 | -3639.312744 |

Figure 5.3.2   Comparison of the performance results for two different models on the main 8 currency pairs

## 5.4   Reinforcement learning

First, we tried different parameters, different input data fragments with different data lengths using AUD_USD_M15 dataset and test the performance of the RL model on the test dataset. The results are given in Figure5.4.1 and the parameter configuration are given in Table 5.4.3. We can observe that for the same dataset, the reward curve for both train dataset and test dataset are very different. Although different parameters have been tried, the model failed to perform as we expected.

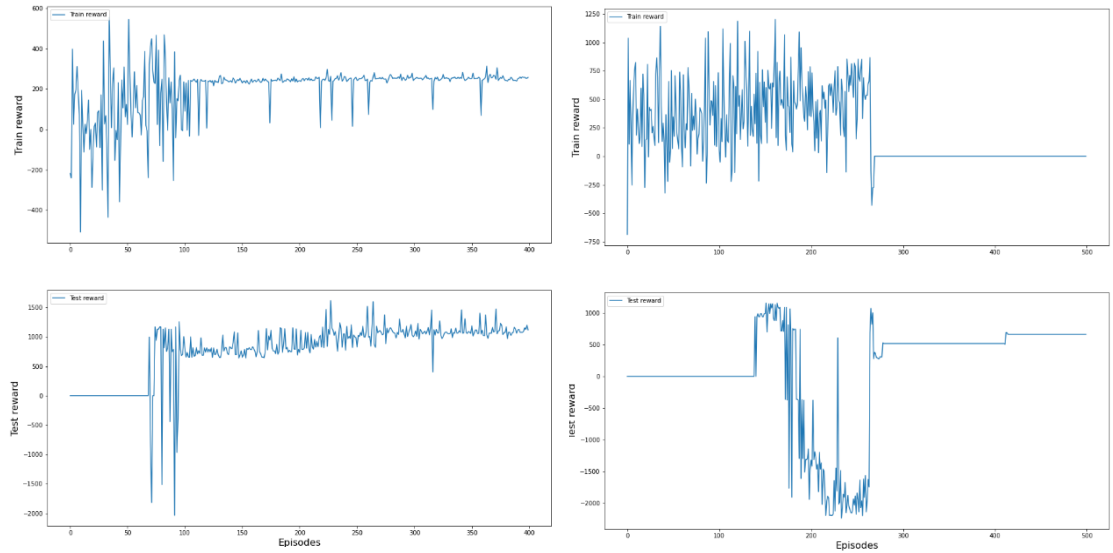

Figure5.4.1. Results under different parameters (left-Model A, right-Model B)

The best model result we achieved is shown in Figure7.4.2. The input data length is 800. We also applied this model on the forecast data generated by the regression model to output a corresponding trading strategy. Because the data is predictive data and the order of the data is disrupted, we cannot test the model on the real data.

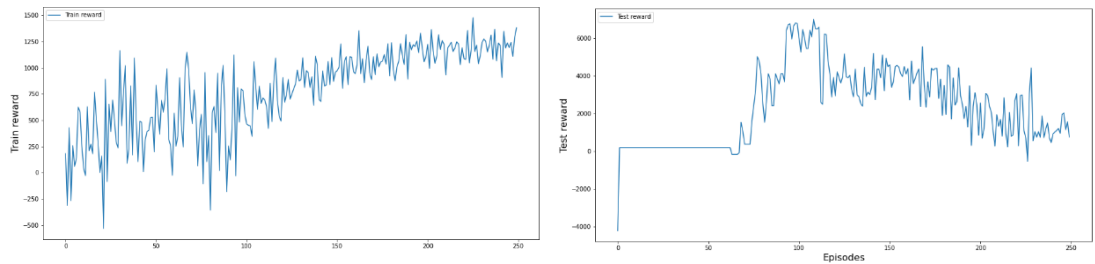

Figure5.4.2. Best result on AUD_USD_M15 dataset (Model C)

Form the figure above, we can conclude that the RL model has a better performance on the predicted data than the test dataset. So, we infer that the RL model may be more suitable for predicting the data that is stationary. Another inference is that the prediction data generated by the regression model contains certain features, which helps the RL model to make better decisions and obtain higher rewards. Details will be discussed in the next section.

|  | $\alpha$ | Data length | Layer size | Episodes | Profits |
|---|---|---|---|---|---|
| Model A | 0.0006 | 800 | 128 | 8000 | -0.05% |
| Model B | 0.0001 | 1600 | 256 | 8000 | -0.03% |
| Model C | 0.0002 | 800 | 128 | 5000 | 1.18% |

Table 5.4.3 The corresponding parameters for 3 models

(Note: $\alpha$ is the learning rate)

In addition to performing experiments using different parameters on the same dataset, we tried to perform the experiments using the same parameters on different datasets. This experiment helped us to find which currency pair will bring us more profits than the others. To ensure the uniformity, all our models will use the same parameters configuration, which is given in the following table.

| $\alpha$ | Input Data length | Layer size | Episodes | Test data length |
|---|---|---|---|---|
| 0.0002 | 1600 | 128 | 12000 | 800 |

Table 5.4.4. Unified parameters configuration for different currency pairs

After the retraining all 7 models, we came to the following result.

| Dataset | Profits | Dataset | Profits |
|---|---|---|---|
| AUD_USD_M15 | 4.92% | GBP_JPY_M15 | 10.15% |
| USD_CAD_M15 | 6.95% | GBP_USD_M15 | 1.14% |
| NZD_USD_M15 | 4.58% | USD_JPY_M15 | 0.68% |
| EUR_USD_M15 | 2.26% | USD_CHF_M15 | 4.075% |

Table 5.4.5. Profits generated by the RL model for different currency pairs

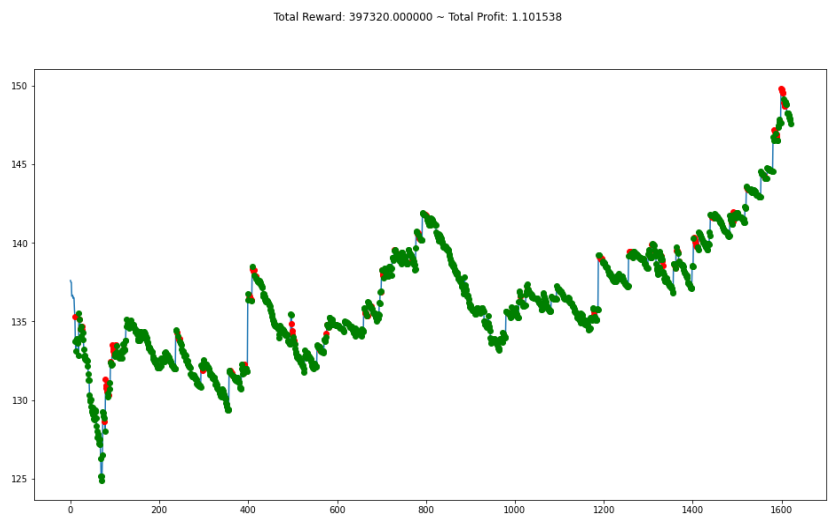

Figure 5.4.6. Environmental rendering for GBP_JPY_M15 (red-sell, green-buy)

Among the eight most traded currencies, we found that our model made the most profit on the GBP_JPY pair, reaching a profit of 10%, but not so good on the other currency pairs.

# 6. DISCUSSION

## 6.1 Clustering + Attention:

Based on the experiment results above, it shows that to predict high price for next 60 mins using last 135 mins sequence length yields the lowest MSE. For next 90 mins high price prediction, using last 120 mins as input sequence length yields the lowest MSE. And for next 120 mins high price prediction, using last 150 mins as input sequence length yields the lowest MSE. Please see below graph for price prediction performance for different window sizes.

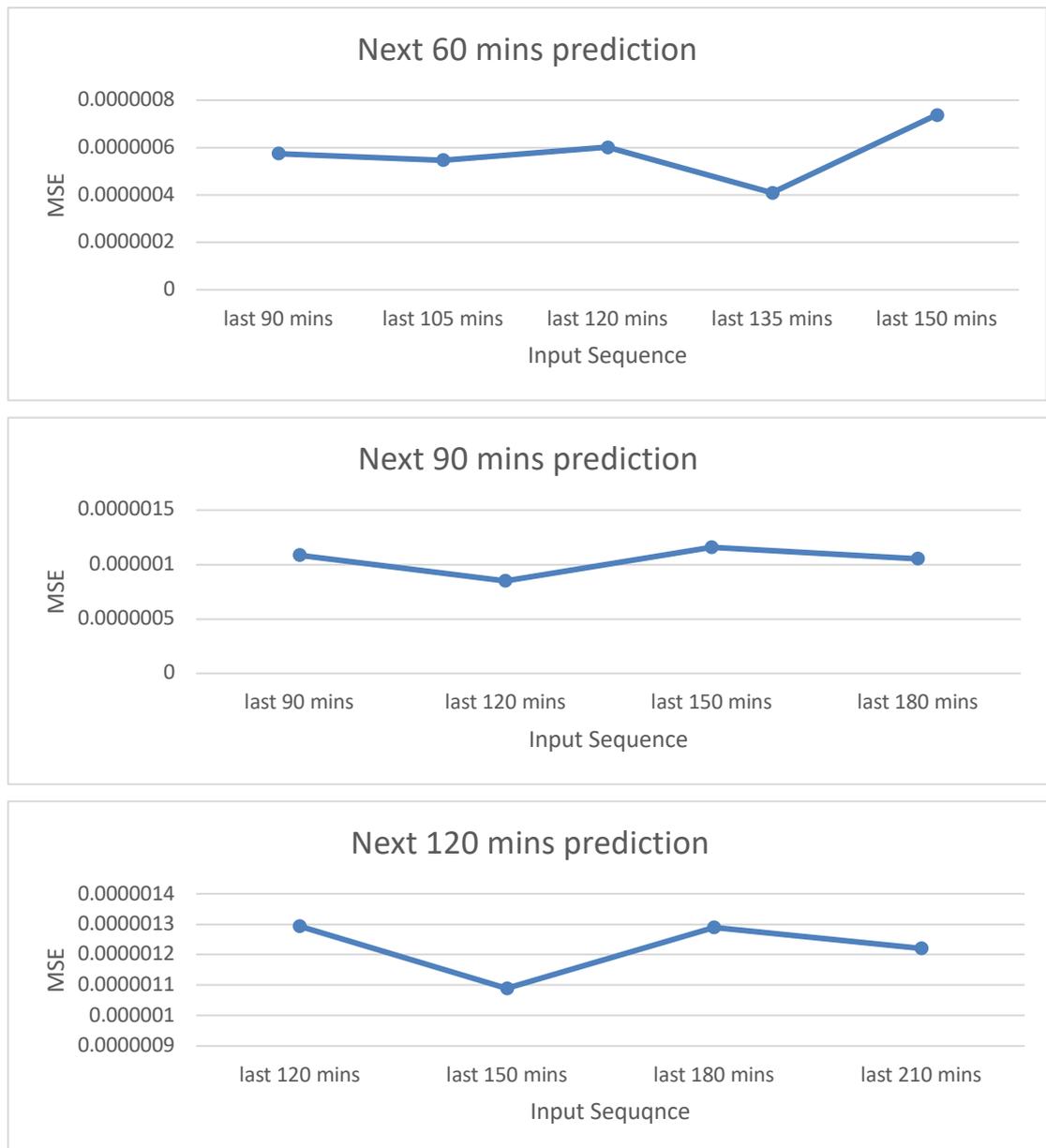

Figure 6.1.1

When input sequence length is fixed at past 90 mins, it works the best with predicting the next 30 mins high price (MSE ~3.65e-07) vs worst performance when predicting the next 105 mins (MSE ~1.28e-06). The reason behind is probably due to smaller pricing fluctuation or movement for shorter period (i.e. price movements within 30 mins on average is smaller than ones in 60 mins). And the smaller price movement will lead to a smaller prediction MSE (see figure 6.1.2). Furthermore, a smaller prediction window size means it is less likely for an instance to be impacted by events such as release of CPI data. For example, if to predict the next 2 days of foreign exchange price movement, it is more likely that we can have some news that could impact the price than if to make prediction for the next 30 mins. Thirdly, with a smaller prediction window size we can have more data i.e. we can have 24 times more data point when predicting next hour than next day's high price. And for most of machine learning models, performance is positively correlated with number of training data. In regard to backtesting P&L, results showing that P&L for smaller window size (i.e. next 30 mins) is limited at ~$3750 (even though it has got smaller MSE) vs ones with large window size (i.e. next 60 mins) ~ $17659 (see below figure 6.1.3). This is because larger price movement usually leads to higher profit margin but with higher risk as well ("lowest capital level" is worse for next 105 mins prediction vs next 30 mins prediction see figure 6.1.4).

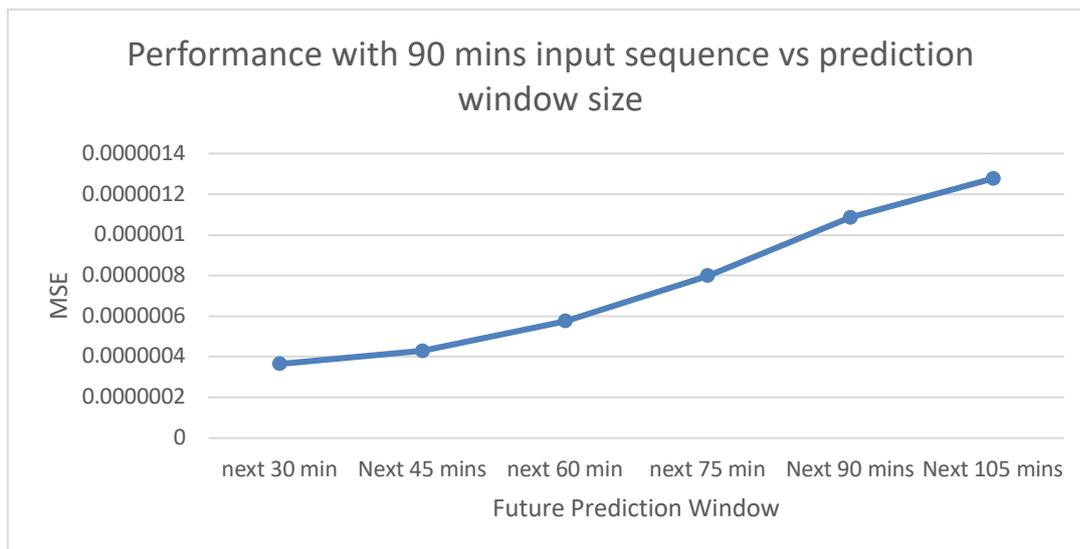

Figure 6.1.2

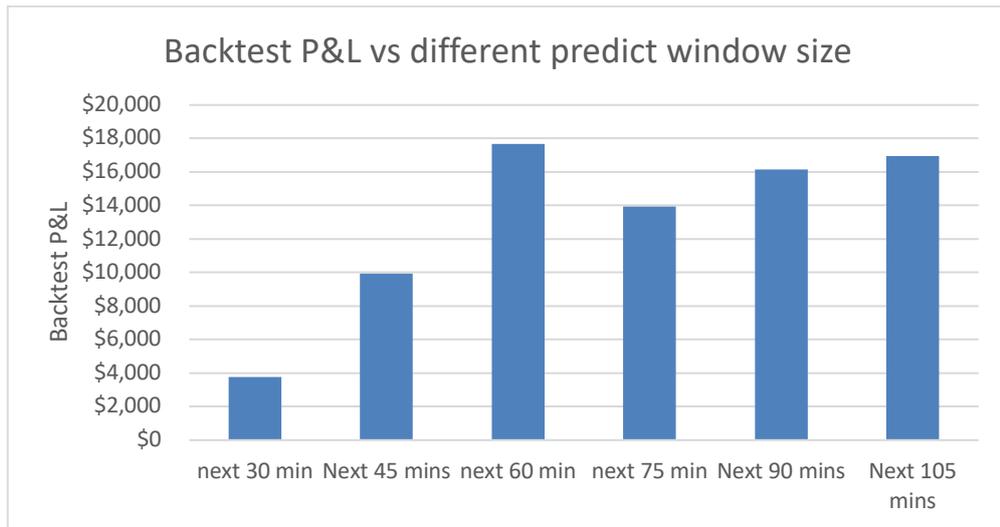

Figure 6.1.3

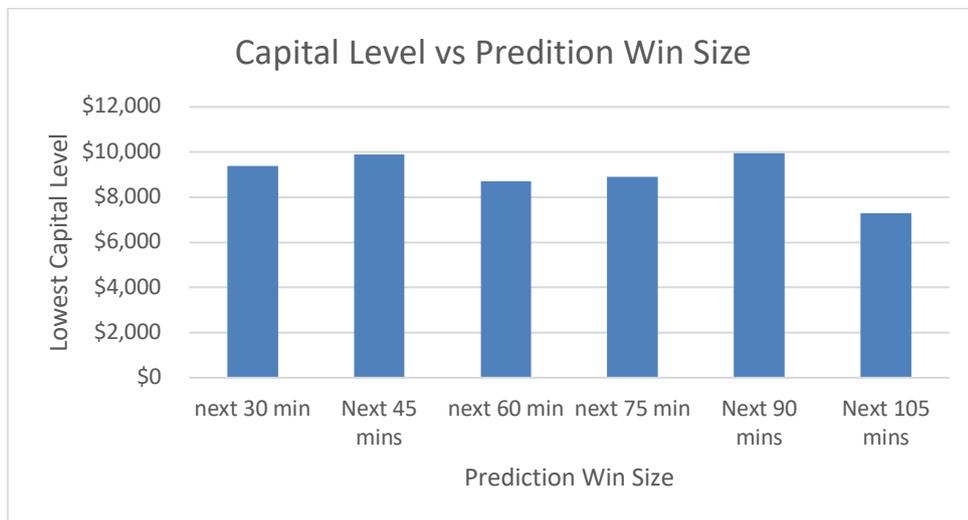

Figure 6.1.4

The model's performances across various currency pairs are different which AUD/USD performs the best and this is probably because AUD/USD is the currency pair we use to train and test our model during every phase of the experiment including feature selection and parameter tuning. It means all the features are selected based on AUD/USD and it might not be the best one for some other currency pairs. And this is especially true for Japanese Yan which is a 2 bps currency pair and results showing our model works poorly against Japanese Yan.

We have also implemented another clustering algorithm – birch. Overall, its optimal performance is very similar to k-means in terms of MSE, RMSE and MAE. One of the key parameters in Birch is cluster threshold. It decides the maximum distance between datapoint for them to be grouped together. In another word, a higher cluster threshold means cluster is more generalised and data are much easier to be grouped together. As per our experiment results, model's performance in terms of both MSE and backtest P&L got much worse when cluster threshold is larger than 0.2.

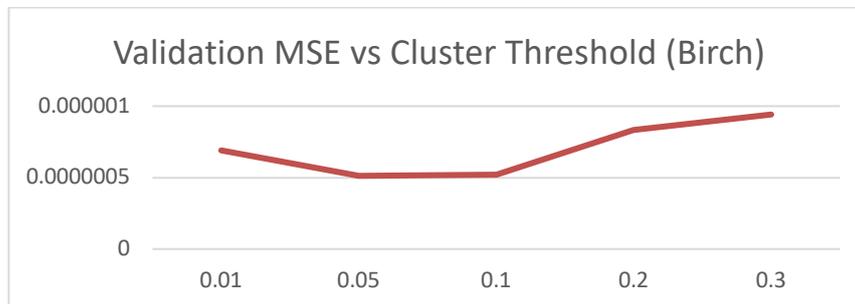

Figure 6.1.5

To find the optimal K value in kmeans clustering, a series of values (from 1 to 18) has been tested. As below graph (fig 6.1.6) shows that test MSE tends to reduce along with higher K. With a small K, the model is not making enough number of sub-groups meaning it tends to allocate data with different pattern into the same clusters. On the other hand, when K is assigned with a large number (i.e. k=16) the performance is stagnated with too many unnecessary clusters. One of the reasons is because we now have less data in every cluster meaning the model is more likely to overfit due to a smaller training data size. Furthermore, for backtesting result between different K value, even though model's MSE (8.54E-07) at K=2 is much worse than K=8 (4.093E-07) the K=2 backtest P&L ($29k) is much higher than K=9 ($14k). This is because the model with k=2 makes more risky trades and it is reflected in capital status: for k=2 its 10k initial capital hit negative during the period meaning its position could have be liquidated by broker. On the other side, the lowest capital level for K=8 is $5.5k based on $10k initial capital shows that a higher K with smaller MSE might not generate the largest profit but will make safer trades for less risk.

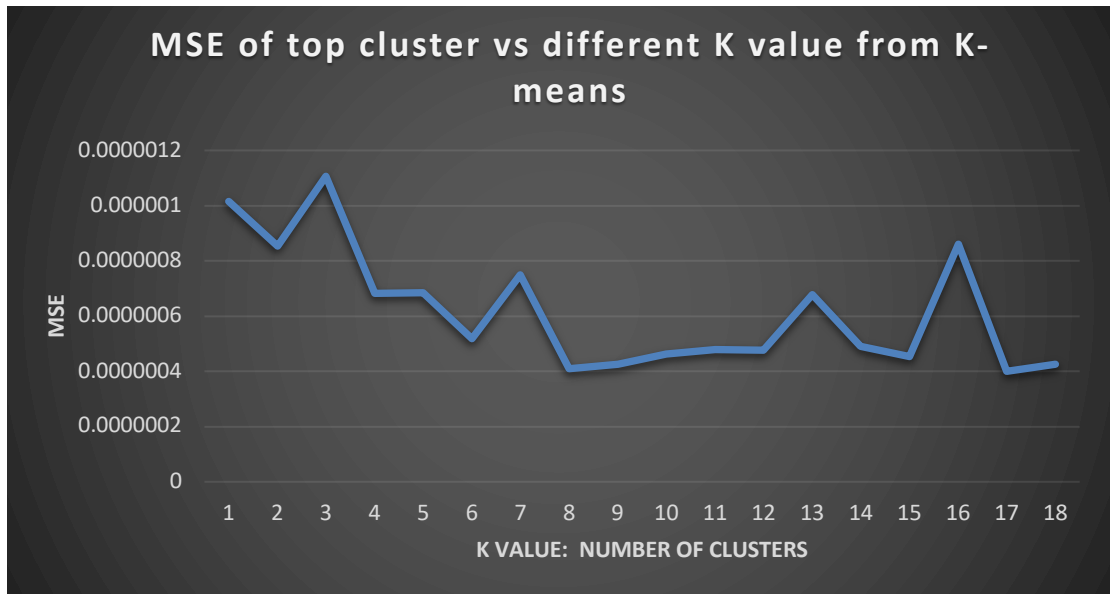

Figure 6.1.6 model's performance using k-means cluster with different K value

Our experiment results also showing that performances are quite different between different clusters (see figure.6.1.7). Among the 9 clusters, the one with the biggest in size (~20% of the total data) has got the lowest test loss. This indicates that the data contained in that cluster can be accurately predictable based on the selected technical indicators. On the other side, we can see that cluster 9 with only 6% of the total data has got a very high test loss. This is an indicator that technical indicators are not good enough to make predictions on those data. We might need other non-technical based features in order to make accurate prediction.

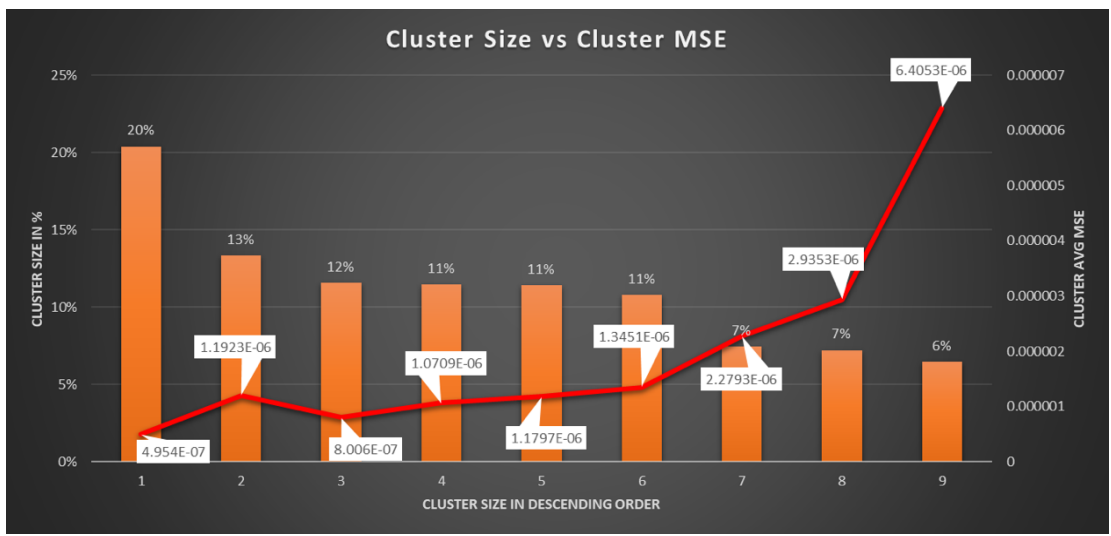

Figure 6.1.7 performance by clusters with k-mean (k=9)

Below is a comparison of performances between our model and one proposed in paper [14]:

| USD/CAD 30 mins | | | |
|---|---|---|---|
|  | MSE | RMSE | MAE |
| **Our Model** | **0.000000636** | **0.000797496** | **0.000549** |
| Hybrid GRU LSTM [14] | 0.00018 | 0.01358 | 0.00998 |
| LSTM | 0.01043 | 0.10211 | 0.10082 |
| GRU | 0.00247 | 0.04965 | 0.03703 |
| SMA | 0.00056 | 0.02374 | 0.01376 |

Table 6.1.8

| EUR/USD 30 mins | | | |
|---|---|---|---|
|  | MSE | RMSE | MAE |
| **Our Model** | **6.077E-07** | **0.000779551** | **0.000485** |
| Hybrid GRU LSTM [14] | 0.00032 | 0.0179 | 0.01233 |
| LSTM | 0.00066 | 0.02573 | 0.01816 |
| GRU | 0.00038 | 0.01955 | 0.01334 |
| SMA | 0.00021 | 0.01469 | 0.01066 |

Table 6.1.9

## 6.2 Transformer Multi Head Attention

Based on our experiment results, time embedding combined with transformer encoder is the method that provides the best result (lowest RMSE) in the task of predicting 1 hour ahead high price for AUD/USD based on 15-minute data, improving the RMSE obtained using transformer encoder with multi-head attention alone.

Positional encoding, which is widely used for NLP tasks, may not be as useful as our input data is already scaled between 0 and 1 so adding another value to each input entry (as we do elementwise addition of positional encoding with input values) will not help the model in making better predictions. Furthermore, the odd and even index concept (sin function applied to even indices i per timestep input, and cosine function applied to odd indices per timestep input) may not be applicable in our context as we are using the features from the same timestep (and their sequence within a timestep is not important).

Using an LSTM layer to capture the sequential information was also not able to provide a good prediction.

Since there are a high number of trainable parameters (over 10k) within the model, a small amount of training data (such as the daily OHLC data) is not sufficient to train the model and we observed underfitting of the data to the model. Similarly, when we attempted event-driven training, validation and testing (filtering by RSI<30 and RSI>70), we observed poor performance as the training data is reduced significantly when filtered by RSI values.

The best performing model includes using convolutional layers for the feed-forward network (as opposed to linear FFN which is used by most transformer architectures). One of the reasons for this discrepancy could be because we are passing 3D output and linear layer would only feed forward information along 1 dimension.

Another observation is that the number of heads along with the dimensions of Key and Value vectors do not make a huge impact to model performance. However, the higher the dimension of n_head, d_k and d_v, the higher the number of computations performed by the model as there are more parameters to tune for the model, which increases the training time. There needs to be a cost vs precision decision made on the best dimensions in such a situation.

The number of encoder heads is also similar where increasing the number of heads increase the performance marginally but also increases the computational complexity of the model so we may need to find more efficient ways to achieve the same accuracy.

The RMSE had a floor at about 0.02, and we attempted to use multi-layer perceptron with GELU activation as a decoder to test if we could make the RMSE lower but the results were worse than using a linear layer for decoder.

Finally, it is important to note that we have only done tests on only a subset hyperparameters within the model. A more thorough hyperparameter tuning setup would involve using a package like Ray Tune and adopting early stopping techniques to determine the best combination of parameters.

### 6.3    Auto Encoder + CNN + LSTM:

It can be learned that from the result, our stacked autoencoder combined with CNN-LSTM model returns RMSE around 0.05 and MSE around 2.5e-03 and from the figure 5.3.1 we can see that the model fitting the actual value within expected. In some of the papers that have been published, the RMSE of CNN-LSTM model can have the magnitude of 10 to the negative power of 4 or more. The reason why the model in this project cannot reaches that accurate can have multiple reasons. The main reason may because of the difference of the dataset. From the result in figure 5.3.2, we are using the same model but different dataset, it can be seen that in some of the currency pairs such as USD/CHF, it has MSE of 7.0e-04 that is lower by an order of magnitude. In the other hand, the currency pair of GBP/JPY has the MSE more than 562. Therefore, the difference of dataset could be the main reason for the difference of the model performance. Secondly, in this project, the features that we used for input into training are the technical analysis indicators which are generated by TA-Lib, this is similar to other papers which was researching about the CNN-LSTM hybrid model. However, the features selection section may be the one of time-costing work due to it decides if there are enough hidden patterns between the input data and the input label. But in this case, we need to be consistent with other models in our group for horizontal comparison and limited of the time, we may not select the best features for our model. Thirdly, the SAE (stacked autoencoder) can be the feature extractor in our model, however, it performs similar to PCA, the data which has gone through the SAE is lossy. Combining with the results in figure 5.3.2, its effect may be interpreted as it will enlarge and focus on the hidden pattern of the data it selected meanwhile it may ignore some of the other key information, this process is random, therefore, sometimes if the SAE removed some key features, will have the situations such as gradient vanishing. Also, it can be seen from the figure 5.3.2, the currency pair of GBP/JPY, the MSE or RMSE is large enough to state that is unreasonable and this is because gradient disappearance. However, by comparing lot of cases, the average performance of SAE + CNN-LSTM model is better than just a CNN-LSTM model in both MSE and profits in back testing. However, due to the randomness of the SAE, in some cases that strictly need stable output, it may be suggested that to have deeper researching on SAE, or for improving the stability of the SAE + CNN-LSTM model, the coding dimension of the SAE can be decided to be larger than it will reduce the information loss of the data, but it may also lead to the decreasing of the effect of features extraction.

### 6.4 Reinforcement Learning:

For the RL model, we cannot expect a huge profit from the current model. In most cases, the algorithm fails to obtain good returns, but there is one thing worth noting that is our RL model has not lose more than 5% of money so far. It proves that our RL algorithm can learn from the historical data and prevent the trading from losing money. However, we still need to improve the model to make it profitable.

According to the results, although the policy gradient algorithm was implemented correctly, the algorithm has not undergone much improvements to increase the learning efficiency, which makes the algorithm require more episodes to complete the learning process. This may cause the RL model remain in the learning process even after thousands of episodes. In other words, due to the high complexity of the policy gradient algorithm, the model did not complete the learning process within the expected episodes. The reason for that might be a small learning rate, which can cause the RL model fall into a local optimal. And we are still finding the possible solution for that problem.

Secondly, we tried to improve the performance of the RL model by modifying the parameters, like the learning rate, running episodes. However, while conducting different experiments, we discovered that for different input data length, we need different parameter configurations. Therefore, we can only find the suitable parameters by trying out several parameter pairs and make adjustments based on the output results. Manual adjustment of parameters greatly increases the difficulty of model optimization and influences the performance of the model.

Finally, we apply the best model to the data predicted by the regression model and generate trading strategies. The training and testing results can be found in Figure 5.4.2. Our model performs well for this data. We infer that the predicted data may have certain features that can improve the performance of the RL model. If these features are applied in the algorithm, then we will be more likely to obtain good results. However, the problem is that if we want to change the input features, we have to re-define the original features that the gym-trading library uses and write our own library. We are still finding the possible solution for this problem.

## 7. LIMITATIONS AND FUTURE WORKS

### 7.1 Clustering + Attention

FX price prediction is a very hot topic and there are a lot of things need to consider in order to develop a workable algorithm. Through this capstone project, one of the models we developed is clustering and attention. Though the model is able to achieve some relatively better performances, there are still a lot of unanswered questions and area for improvements. Firstly, how do we make optimal feature selections for clustering? How do we know whether the way cluster splits the data is what we expected i.e. how to split technically predictable data from the rest. This is an important area that can really bring the performance or use of the model to next level as prediction performance especially profit generation ability is highly dependent on clustering performance. For example, we have done some experiments with different technical indicators as input features. One interesting finding is that some technical indicators can improve test losses (i.e. MSE or MAE). But when it comes to backtesting, that also led to some bad losses (i.e. -200% in profit). The reason is our model can make good prediction on most of the data using the selected technical indicators. However, model perform extremely badly on 4~5 out of ~2000 testing instances. These 4 to 5 instances do not really impact our overall MSE or MAE losses but impose a material impact on backtesting. And to solve the problem, we realise that the issue is with clustering. Since some technical indicators are good at prediction but is not good for use during clustering. Therefore, we decided to use two different sets of technical indicators for prediction and clustering. Since then, we are able to eliminate most of those losing trades and end up with a profit. Therefore in order to make the model work well especially in backtesting, the performance of clustering is extremely important. A lot of research needs to be done to work out what technical indicators are better to use in order to allocate data to different groups the way people wanted.

In addition, we only use one prediction algorithm (attention) for every cluster and it generates mixed results (great for some clusters and much worse for others) meaning that number of predictions that our model can make is limited (since we can only focus on clusters that our model performs well). So to develop different prediction algorithms and different feature sets for different clusters (i.e. even to use NLP related techniques to obtain non-numerical information as input feature for certain clusters) can further leverage the performance of the model. Also, not to only focus on oversold event, but to further develop the model so that it can work on other events as well which can create more trading opportunities.

Thirdly, we have not utilised the use of minimum price prediction. The minimum price prediction can play an important role in trading strategy i.e. it can be used to assess trading risks. This is especially true if our model can extend its use beyond oversold condition.

Fourthly, the model has performed poorly against JPY related currency pair. Meaning the model cannot be applied to some other financial instruments for price prediction directly without fine-tuning or re-selection of features. Therefore, some functions or automation can be added to the model so that it can auto-select important features when dealing with a new financial instrument.

Furthermore, unlike other machine learning task financial market keeps evolving over time meaning making prediction based on old historical data might lead to incorrect results. Therefore, the model needs to be updated with the most recent trend in FX market in order to maintain its performance (maybe give extra weights to more recent data during training). In some cases, model might need to abandon some old historical data and focus on more recent ones in order to make more accurate predictions. So how to keep the model updated with latest market behaviour or evolve along with the financial market is another important area for future improvement.

Lastly, the model has not been tested or deployed to a live trading environment. Meaning even though we are confident that the model can generate some level of positive income, it cannot be used directly without any modifications or further development in order to engage live trading. As a future improvement, we can find a way to enable the model so that it can feed on live market data and make decisions then execute trades automatically. This is one of the ultimate goals for developing the model.

Based on timing and other resource constraints, there are a lot of questions left unanswered. And we believe that this structure has the potential to be one of the paths to solve FX pricing prediction problem. Yet there are a lot more tasks need to be done to make it work with people's expectations.

### 7.2  Transformer Multi Head Attention

There are several limitations to the Transformer encoder model which will prevent our solution from being able to predict accurately and be deployed into industry or a production environment. Firstly, the model needs a large dataset to train on to achieve an acceptable level of RMSE. The RMSE for our model is still very high (and unstable) and it cannot be used for real time forex trading as the error in prediction is not reliable enough to be able to make enough profit to cover fees. It is also slow to train, which makes it unfit to deploy in a production or live trading environment. The results on different currencies indicate that the model does not generalise well for all currency pairs. Therefore more work needs to be done to make it to prevent overfitting to one particular currency pair, i.e. AUD/USD in our experiment.

We also need to consider the fact that transformer-based models may not be suited for price forecasting tasks in general and that the architecture is more suited to NLP domain where it is widely used with a lot of success. Like the modification to the embedding layer, we may need to make further modifications to the architecture in order to get better prediction results. This can be done using an iterative process of trial and error.

In terms of project constraints, we are limited in terms of time, economic and computational resources. We might be able to achieve better results if we dedicated more time (more than the 12-week semester) or were able to work on the project full time as an example. If we had better computational resources such as TPUs available to perform testing and training, that may help in developing a better model and faster training. In terms of financial constraints, if we had the ability to hire a SME or some professional developers then that may also help in achieving a better result.

In terms of future work for our model, we could extend the model to have a decoder and masking layers similar to an NLP transformer. However, it remains to be seen if that will improve the performance of the model as we are not doing a language translation task, so having more layers of multi-head attention may not necessarily improve the model.

Another avenue to extend the work we have done is to train the model to generalise better on a variety of currency pairs or stock prices/stock indices. We may try to extend the model to try to predict cryptocurrency exchange rates as an example, although this may prove to be a challenge given the high volatility of cryptocurrency values (a lot more volatile than financial markets which are also volatile to begin with).

### 7.3   Auto Encoder + CNN + LSTM:

In regard to limitation, the reason why the forex or stock price is predictable is because most of the case in trading market is artificially controlled and group behaviour. Therefore, the raw dataset which only includes high, low, close and open are not enough to have perfect prediction. Some of the data such as volume, bids and offers, the data in course of sales may contain more information of human behaviour. As we mention above in the discussion, the dataset may have a large impact on performance of the model.

Second limitation would be the time constraints. Since the SAE is a neural network based, thus it is equivalent to have 2 neural networks trained at the same time. In this case, we have lots of hyperparameters to be tuned based on architecture of the SAE and CNN-LSTM model. Another time-costing section is the selection of features, since there are many technical indicators for forex market, we do not have enough time to try which indicators are helpful and contain more hidden information. Furthermore, some technical indicators are binomial variable since the analysis pattern do not occur frequently. However, the features selection could be an important factor to affect the performance of the model. For further research in this project, we can have much more time for feature engineering and may not only focus on technical indicators only.

In regard to future improvement, the training process in this project is using the data of 8 last time point to predict one target value, the result shows that our models have performed within our expectation to some extent. However, in real world trading, especially in other financial market such as stock market, it may not be helpful for just predict the value of one future time point. As future works, we can develop the rolling forecasting that to predict a series of values covers next period.

In real world trading, a successful trader would be more wary about the policy and long-term trend. Since technical analysis is more often used for short-term trading and it has significant disadvantage. An important reason is that the technical analysis is not very robust to the future data, even though the model makes substantial profits, it may lose all the gains while there is a reversion in financial market. To build a comprehensive machine trading system, we can expand the feature engineering and the way we forecast. We can include the news, the policy of the trade or even the fundamental of company data if we are applying the model in stock market. Thus, the nature language processing and similar techniques can be used in the future.

Also, if time is not constraint, we can develop more trading strategies such as buy long or contract for difference that making profits from downward trend, hedging for reducing trading risk. In the future, we can try different combination of trading methods to see what strategies can make more profits by back testing and real-world trading.

## 7.4 Reinforcement Learning

The limitation of the RL model lies in the complexity of the algorithm. To be specific, the parameter tuning process is complicated and training process is time-consuming.

Firstly, for the basic policy gradient algorithm, the most important factor affecting the performance of the model is the parameter configuration. From the results of the experiments, it was found that the RL model is more sensitive to parameters than we thought. Small adjustments to the parameters will greatly affect the output results. So, for the RL model it is not as easy to adjust parameters as other deep learning models and it has a worse robustness.

This property makes it challenging to tune the parameters and we are currently finding improving methods that we can implement in the future. One of the algorithms we have found is called proximal policy optimization (PPO). According to an article written by Jonathan Hui [36], the PPO algorithm helps the model to achieve an increasing reward after each episode and is more robust. Another method is called divide and conquer which is introduced by Dibya Ghosh and his colleges [37]. It is a combined model. In the article, they separate complex tasks into a set of local tasks, and each of which can be used to learn a separate policy. These separate policies are constrained against each other to arrive at a single, globally coherent solution, which can then be used to solve the task from any initial state. This algorithm can greatly improve the performance of the RL model, but it also increases the complexity of the computation.

Secondly, the training process of the RL algorithm is time-consuming. For some projects without a time limitation or are not strict to time, like learn how to play games or fly a helicopter, it does not matter for how long the RL algorithm runs. Because the desired results will still be obtained eventually. But for the financial market, the price and environment change every second, and a model that needs to run for a whole day is not advisable. To solve that problem, we believe that combining LSTM with the RL model is a desirable solution. The LSTM algorithm will collect the features that have a relatively large impact on the future reward, and forget some of the less influential factors, thereby increasing the learning efficiency of the model.

Another limitation for the RL model is the gym-trading library. First, there is no action 'hold' in this library, which means that we are constantly 'buying' or 'selling'.

Although the author of the library claimed that the reason for not holding is that they just want to improve the learning time and an action like 'hold' will be barely used by a well-trained agent because it does not want to miss a single penny [38]. But we cannot perform the tests to observe the impact of 'hold' on our model. In addition, from the conclusion, the selection of data and the features contained in the data are also very important. In this library, it is hard for us to change the input feature because it is fixed. In the future, we may try to derive some important features from the historical data, and then select the data based on the features other than RSI and use the selected data to training the model. To do so, we may need to write an applicable reinforcement learning environment and re-define the input feature.

In summary, our future improvement will mainly focus on reducing complexity of tuning parameters and improving the robustness of the model. In addition, we will try to combine different models to improve the learning efficiency and shorten the learning time. Finally, we will also try to write an environment suitable for foreign exchange market so that we can change the input of the model based on the needs.

## 8. CONCLUSION

In conclusion, we have implemented four deep learning models with very different basic ideas and architectures. Some of the models have been able to generate good performance comparing with others. FX market is unique and complicated in many ways which makes its price prediction task be very different from other machine learning tasks such as image recognition or language generation. To tackle the problem in the right direction, domain knowledge such as understanding of how FX market works and experience in trading is important. And we hope the methods presented in the report can provide some inspirations and hints for those in the future.

*Conference on Data Mining Workshops (ICDMW)* (pp. 385-391). IEEE.